\documentclass[12pt,a4paper]{article}

\usepackage{latexsym,amssymb}
\usepackage{graphicx,color}
\usepackage{amsmath}
\newcommand\sect[1]{\section{#1}\setcounter{equation}0}

\newcommand\no{\nonumber\\{}}

\newcommand\eqnb{\begin{eqnarray}}
\newcommand\eqne{\end{eqnarray}}

\newcommand{\mf}{\mathfrak}

\newcommand{\C}{\ensuremath{\mathbb{C}}}

\newcommand{\Z}{\ensuremath{\mathbb{Z}}}

\usepackage{latexsym}

\newcommand\void[1]{}   

\newcommand{\al}{\alpha}
\newcommand{\be}{\beta}

\newcommand{\de}{\delta}
\newcommand{\hal}{\hat{\alpha}}

\newcommand{\hmu}{\hat{\mu}}
\newcommand{\hDe}{\hat{\Delta}}
\newcommand{\hrho}{\hat{\rho}}
\newcommand{\vs}{\vspace*}
\newcommand{\vsv}{\vspace*{5mm}}
\newcommand{\hs}{\hspace*}

\newcommand{\Tr}{{\mathrm{Tr}}}
\newcommand{\ii}{{\mathrm{i}}}
\newcommand{\sign}{{\mathrm{sign}}}
\newcommand{\bi}{\beta^{(1)}}
\newcommand{\bii}{\beta^{(2)}}
\newcommand{\la}{\lambda}

\newtheorem{theorem}{Theorem}

\newtheorem{lemma}{Lemma}

\begin{document}

{\center{{\huge WZNW Strings, Unitarity and \\ Anti de Sitter spaces}\vspace{15mm}\\
Jonas Bj\" ornsson\footnote{j.bjornsson@damtp.cam.ac.uk} and Stephen 
Hwang\footnote{stephen.hwang@kau.se}$^,$\footnote{stephen.hwang@lnu.se}\\
\vspace{.5cm}
$\mathrm{{}^1}$Department of Applied Mathematics and Theoretical Physics\\
Cambridge University\\
Wilberforce Road\\
Cambridge CB3 0WA, United Kingdom\\
\vspace{.5cm}
$\mathrm{{}^2}$Department of Physics\\
Karlstad University \\
SE-651 88 Karlstad, Sweden\\
\vspace{.5cm}
$\mathrm{{}^3}$Department of Physics\\
Linn\ae us University \\
SE-391 82 Kalmar, Sweden}\vspace{15mm} \\}


\begin{abstract}
We investigate the unitarity of strings on non-trivial space-time backgrounds based on gauged WZNW models involving $SO(n,2)$, gauging an $SO(n,1)$ subgroup. As geometric coset spaces, these are Anti de Sitter spaces. Our present models are more complicated than the ones considered previously, for example those connected to Hermitian symmetric spaces. In the present case, the time-like field component is not a center of the maximal compact subalgebra, leading to several additional complications. Assuming discrete representations, it will only be possible to consider $n$ even, resulting in odd dimensional spaces. We will prove that such models are free of ghosts for a class of discrete representations. 
\end{abstract}


\sect{Introduction}
In a series of papers \cite{Bjornsson:2007ha}--\cite{Bjornsson:2009dp} we have studied string propagating on coset manifolds constructed using WZNW models with a non-compact group $G$ gauging $H'$. $H'$ is constructed from the maximal compact subgroup $H$ as $H= Z(H)\times H'$ where $Z(H)$ is the one-dimensional center of $H$. The spaces $G/H$ are non-compact Hermitian spaces, which have all been classified, see \cite{Helgason}. The unitarity of these models relies on BRST construction \cite{Karabali:1989dk,Rhedin:1995um} (see also \cite{FigueroaO'Farrill:1995pv}) as the GKO construction \cite{Goddard:1984vk} yields a non-unitarity spectrum \cite{Bjornsson:2007ha}. This is the first case where the two approaches yield different results.

These models have been studied extensively, see \cite{Balog:1988jb}--\cite{Henningson:1991jc} for early work, More recently, the progress made to understand the non-perturbative properties of string theory leading to M-theory \cite{Townsend:1995kk,Witten:1995ex} and D-branes \cite{Polchinski:1995mt}, the non-compact backgrounds associated with Anti de Sitter (AdS) spaces became of central interest \cite{Maldacena:1997re}. 

The progress since the original proposal of the AdS/CFT-conjecture has been substantial (see \cite{Arutyunov:2009ga} for a recent review). One of the most important steps is the connection to integrable models. In \cite{Bena:2003wd} it was shown that strings moving on AdS${}_{5} \times$S${}^5$, based on the action suggested in \cite{Metsaev:1998it}, has an infinite number of conserved charges and are, therefore, classically integrable. An alternative approach to the one in \cite{Metsaev:1998it} is the covariant approach using pure spinors \cite{Berkovits:2000fe}. 

The form of the Hamiltonian in light-cone gauge was derived in \cite{Arutyunov:2004yx,Frolov:2006cc}, and has a complicated structure. One can study, however, a particular limit in which the dynamics simplifies. This is the limit where the light-cone momentum is taken to infinity while holding the physical string tension fixed. The world-sheet theory then has a massive spectrum and a well defined notion of asymptotic states. In this limit, quantum integrability implies that there is no particle production and that multi-particle scattering amplitude should factorize into two-body scattering ones. Considerable progress has been made assuming quantum integrability of the model. One has determined the spectrum of bound states of the model \cite{Dorey:2006dq}, conjectured the dispersion relation \cite{Beisert:2004hm} and determined the matrix structure of the S-matrix \cite{Beisert:2005tm}--\cite{Arutyunov:2007tc}. Furthermore, one has found an asymptotic form of the dressing factor of the S-matrix \cite{Beisert:2006ib} and conjectured the exact form \cite{Beisert:2006ez}. The progress has been for large light-cone momentum, but there has also been progress for finite light-cone momentum. One has fairly recently conjectured that the so-called Y-system \cite{Zamolodchikov:1991vg,Kuniba:1994nn} encode the finite size string spectrum \cite{Gromov:2009tv}.

In this paper we will study string theories on backgrounds constructed as gauged WZNW models, where $G=SO(n,2)$ and one gauges $H'=SO(n,1)$. More precisely, we will study the unitarity problem for such string theories. As geometric coset spaces, these are Anti de Sitter spaces of dimension $n+1$. It should be remarked, however, that the string background we consider is the one where one gauges an adjoint action of the subgroup yielding a space $G/\mathrm{Ad}(H')$, which is not the same space as the geometric coset space. We will be able to construct a consistent gauging when $n=2p$ and prove that these models give a unitary spectrum for a class of discrete representations. This provides the first example of a unitary string theory beyond the class of models studied previously \cite{Bjornsson:2007ha}--\cite{Bjornsson:2009dp}, associated with Hermitean symmetric spaces. One important difference compared to the previous class of models is the absence of the central element $Z(H)$. This element plays the role of time in the previous models. In the present class of models, time is embedded in a more complicated fashion. This will lead to several additional complications. One difficulty will be to find a consistent embedding of the subalgebra. This is the reason why there is no regular embedding for the even dimensional case. Another difficulty will lie in the proof of unitarity, which will turn out to be more involved. 

The models treated here are easily generalized to world-sheet supersymmetric ones. The proof of unitarity is readily generalizable to this case, as is briefly discussed in the paper. However, space-time supersymmetry is not achieved and it is not known how to implement it in our approach. Although our model is distinctly different from \cite{Metsaev:1998it}, there might be some common features. One can reduce the action, using the Pohlmeyer reduction \cite{Pohlmeyer:1975nb}--\cite{Pohlmeyer:1979ch}, to a gauged WZNW model with world-sheet fermions and a perturbation \cite{Grigoriev:2007bu}--\cite{Hoare:2009fs}. This gauged WZNW model is gauge fixed, and the relevant group is $SO\left(4,1\right)$ with a gauging of $SO\left(4\right)$. 

The paper is organized as follows. Section two is devoted to the formalism, the third section discusses how $\mf{so}(2p,1)$ is realized as a subalgebra in $\mf{so}(2p,2)$ and how it is gauged. The fourth section proves that the state-space is unitary for the choice of representations. In this section we also discuss the generalization to the world-sheet symmetric model. In the fifth section we discuss the problems in formulating and, therefore, proving necessary conditions for the coset to yield a unitary spectrum. The last section contains some concluding remarks. In an appendix we discuss some properties of $\mf{so}(4,2)$ and the subalgebra $\mf{so}(4,1)$.


\sect{Formalism}

The conventions and definitions used in this paper are based on \cite{Fuchs:1997jv}. Denote by $\Delta$ all roots, $\Delta^{+/-}$ the positive/negative roots, $\Delta_s$ the simple roots, $\Delta_c$ the compact roots, $\Delta_c^+=\Delta_c\cap\Delta^+$ the compact positive roots, $\Delta_n$ the non-compact roots and $\Delta_n^+$ the positive non-compact roots. We take the long roots to have length $\sqrt 2$. Let $\alpha\in \Delta$ and define the coroot by $\al^\vee=2\left(\al,\al\right)^{-1}\al$. Let $\alpha^{(i)}\in \Delta^+$ denote the simple roots. 

The Cartan-Weyl basis of $\mf{g}^{\mathbb{C}}$ is
\eqnb
\left[H^i,H^j\right]
      &=&
          0,
      \no
\left[H^i,E^{\al}\right]
      &=&
          {\al^{i}}E^{\al},
      \no
\left[E^{\al},E^{\be}\right]
      &=&
          e_{\al,\be}E^{\al+\be}+\delta_{\al+\be,0}\sum_{i=1}^{r_{\mf{g}}}\al^{\vee}_iH^i,
\eqne
where  $e_{\al,\be}\neq 0$ if $\al+\be\in\Delta$, $\alpha^i$ are components in the Dynkin basis, $\al^\vee_ i=\left(\al^\vee,\Lambda_{(i)}\right)$. Here $\{\Lambda_{(i)}\}$, $i=1,\ldots , r_{\mf{g}}$, is a basis of the weight space.

The Cartan-Weyl basis can be extended to the affine Lie algebra $\hat{\mf{g}}^{\mathbb{C}}$,
\eqnb
\left[H^i_m,H^j_n\right]
      &=&
          mk{G}^{(\mf{g})ij}\delta_{m+n,0},
      \no
\left[H^i_m,E^{\al}_n\right]
      &=&
          \al^iE^{\al}_{m+n},
      \no
\left[E^{\al}_m,E^{\be}_n\right]
      &=&
          e_{\al,\be}E^{\al+\be}_{m+n}+\delta_{\al+\be,0}\left(\sum_{i=1}^{r_{\mf{g}}}\al^{\vee}_iH^i_{m+n} 
          + \frac{2}{(\al,\al)}mk\delta_{m+n,0}\right),
\label{affalgebra}
\eqne
where $G^{(\mf{g})ij}=\left(\al^{(i)\vee},\al^{(j)^\vee}\right)$ is the metric on the weight space with inverse ${G}^{(\mf{g})}_{ij}=\left(\Lambda_{(i)},\Lambda_{(j)}\right)$ and $k$ is the level. We denote by $\hat{\Delta}$ the affine roots and by $\left| 0;\mu\right>$ a highest weight state of a $\hat{\mf{g}}^{\C}$ module with weight $\hat{\mu}=\left(\mu,k,0\right)$. It satisfies
\eqnb
{J}^{\hat{\mf{g}}}_{+}\left| 0;\mu\right> 
      &=&
          0,      \\
H^i_0\left| 0;\mu\right>
      &=& 
          \mu^i\left| 0;\mu\right>,
\eqne
where 
$J^{\hat{\mf{g}}}_{+}=\{H^i_m,~E^\al_n,~E^{-\al}_m\}$ for $m>0, n\geq 0$ for $\al\in \Delta^+$. We also define $J^{\hat{\mf{g}}}_{-}=\{H^i_{-m},~E^{-\al}_{-n},~E^\al_{-m}\}$ for $m>0, n\geq 0$ for $\al\in \Delta^+$. The irreducible highest weight $\hat{\mf{g}}^{\C}$ module that is defined by acting with $J^{\hat{\mf{g}}}_{-}$ on the highest weight state is denoted by ${\cal H}^{\hat{\mf{g}}}_{\hat\mu}$.

The generators are defined to have the Hermite conjugation properties $(J^{\hat{\mf{g}}}_{+})^\dagger =\pm J^{\hat{\mf{g}}}_{-}$ w.r.t.\ $\hat{\mf{g}}$, where the minus sign appears for $\alpha\in{\Delta}_n$ and the plus sign otherwise. The generators of the Cartan subalgebra are Hermitian. The norm of $\left| 0;\mu\right>$ is defined to be one. Norms of other states $\left|s'\right>\equiv J^{\hat{\mf{g}}}_{-}\left|s\right>$ are then defined iteratively by $\left<s'|s'\right>=\left<s\right|(J^{\hat{\mf{g}}}_{-})^\dagger J^{\hat{\mf{g}}}_{-}\left|s\right>$.

A weight $\hmu$ is said to be dominant if $(\hal, \hmu)\geq 0$, $\hal\in \hDe_s$ and  antidominant if $(\hal, \hmu+ \hrho)< 0$, $\hal\in \hDe_s$. Here $\rho=\frac{1}{2} \sum_{\al\in\Delta^+}\al$ and $\hrho$ is the affine extension of this, $\hrho=(\rho, g^\vee, 0)$, where $g^{\vee}$ is the dual Coxeter number. The dominant and antidominant weights are said to be integral if they have integer components. Integral dominant highest weight representations are often called integrable. Dominant affine weights require $k\geq \left(\theta,\mu\right)$ and $\left(\mu,\alpha^{(i)\vee}\right)\geq0$. Antidominant affine weights require $k + g^\vee<0$, $k<\left(\theta,\mu\right)-1$ and $\left(\mu+\rho,\alpha^{(i)\vee}\right)<0$. The Shapovalov-Kac-Kazhdan determinant \cite{Sapovalov, Kac:1979fz}, implies an important property of antidominant weights, namely that the corresponding highest weight Verma modules are irreducible. This follows since the determinant is always non-zero.


\sect{Gauging $\mathfrak{so}(2p,1)$}\label{sec:emb}
The construction we will be considering here is the coset $SO(2p,2)/SO(2p,1)$. The relevant complex algebra is $D_{p+1}$ for the real form $\mf{so}(2p,2)$, which is of rank $p+1$. Similarly, $\mf{so}(2p,1)$ is a real form of $B_p$ and is of rank $p$. $D_{p+1}$ has a dual Coxeter number $g^\vee=2p$. $\mf{so}(2p,2)$ has $4p$ non-compact generators and $p(2p-1)+1$ compact ones. $B_p$ has a dual Coxeter number $g^\vee=2p-1$. $\mf{so}(2p,1)$ has $2p$ non-compact generators and $p(2p-1)$ compact ones. We will, for convinience, call a root $\alpha$ (non-) compact if the corresponding generator $E^\alpha$ is (non-) compact. 

Let us collect some properties for $D_{p+1}$ w.r.t. the real form $\mf{so}(2p,2)$. 
\begin{lemma}
(i) One can choose a basis such that the simple roots $\alpha^{(p)}$ and $\alpha^{(p+1)}$ 
are non-compact, while the rest of the  positive non-compact roots are 
\eqnb
\alpha^{\{r\}}&=&\alpha^{(p)}+\sum_{i=1}^{r-1}\alpha^{(p-i)}\no
\alpha^{\prime\{r\}} &=&\alpha^{(p+1)}+\sum_{i=1}^{r-1}\alpha^{(p-i)},
\label{D-non-compact}
\eqne
for $r=1, \ldots , p-1$.\\
(ii) $\al^p-\al^{p+1}=0$ for $\al\in\Delta_c$, where $\al^i=\left(\al,\al^{(i)\vee}\right)$.
\label{lemma D-roots}
\end{lemma}
{\bf Proof.} The above statements are easily proven using an orthogonal basis $\left\{e_i\right\}_{i=1}^{p+1}$(see eg. \cite{Fuchs:1997jv}). Then the positive roots can be written
\eqnb
\al=\left\{e_i\pm e_j\right\}_{1\leq i<j\leq p+1},
\label{roots-orthogonal}
\eqne
and the simple roots
\eqnb
\al^{(i)}&=&\left\{e_i- e_{i+1}\right\}_{1\leq i\leq p}\no
\al^{(p+1)}&=&\left\{e_p+ e_{p+1}\right\}.
\label{simpleroots-orthogonal}
\eqne

To prove assertion $(i)$ we should show that by choosing $\alpha^{(p)}$ and $\alpha^{(p+1)}$ non-compact while the rest of the simple roots are compact, we will get the correct number of compact and non-compact roots. Using the orthogonal basis above, we first see that the roots defined in eq.~(\ref{D-non-compact}) are indeed roots. Furthermore, they are obviously non-compact. This gives us altogether $2p$ non-compact positive roots, which is the correct number. The roots that are left are of the form $\al=\left\{e_i\pm e_j\right\}_{1\leq i<j\leq p}$. These roots can never involve an odd number of non-compact simple roots, as such roots would always contain $e_{p+1}$, so the roots not of the form eq.~(\ref{D-non-compact}) are compact. This means that all compact roots lie in the $p$-dimensional space spanned by $\left\{e_i\right\}_{i=1}^{p}$.

To prove  $(ii)$ we simply use 
\eqnb
\al^p-\al^{p+1}=\left(\al,\al^{(p)\vee}-\al^{(p+1)\vee}\right)=-2\al\cdot e_{p+1},
\eqne
which, by using that the compact roots are
$\al=\left\{e_i\pm e_j\right\}_{1\leq i<j\leq p}$, gives the assertion.
$\Box$

It should be noted that the basis considered above is not the same as in \cite{Bjornsson:2007ha,Jakobsen:1983,Enright:1983}. There the basis was chosen such that there was a unique simple non-compact root. The highest root, $\theta_{D_{p+1}}$, in the latter basis is non-compact, while in the present one it is compact. Somewhat surprisingly, the present basis does not admit unitary highest weight representations (except for $\mf{so}(2,2)$), contrary to the previous basis. A discussion of this and other properties for the case $\mf{so}(4,2)$ is given in the appendix. The reason for the present choice of basis is that it allows a regular embedding of $\mf{so}(2p,1)$ into $\mf{so}(2p,2))$ i.e. one that preserves the triangular decomposition of the algebra. The former basis does not allow a regular embedding. This is shown for the case $\mf{so}(4,2)$ in the appendix and since $\mf{so}(4,2)$ is a subalgebra for the higher dimensional cases, it holds in general.

\begin{figure}[h!]

\caption{Dynkin Diagrams for $B_p$ and $D_{p+1}$}

\begin{center}
\begin{tabular}{|c|c|}

\hline

\begin{picture}(160,90)(-20,-30)


\put(0,40){Dynkin diagram of $B_p$:}

\multiput(0,0)(30,0){5}{\circle*{4}}

\put(0,0){\line(1,0){30}}
\put(60,0){\line(1,0){30}}

\multiput(90,-2)(0,4){2}{\line(1,0){30}}

\multiput(30,0)(7.5,0){5}{\circle*{1}}

\put(100,-10){\line(1,1){10}}
\put(100,10){\line(1,-1){10}}

\put(-2,-15){\footnotesize{1}}
\put(28,-15){\footnotesize{2}}
\put(53,-15){\footnotesize{p--2}}
\put(83,-15){\footnotesize{p--1}}
\put(118,-15){\footnotesize{p}}

\end{picture}

&
 
\begin{picture}(160,90)(-20,-30)


\put(0,40){Dynkin diagram of $D_{p+1}$:}

\multiput(0,0)(30,0){4}{\circle*{4}}
\multiput(110,-20)(0,40){2}{\circle*{4}}

\put(0,0){\line(1,0){30}}
\put(60,0){\line(1,0){30}}
\put(90,0){\line(1,1){20}}
\put(90,0){\line(1,-1){20}}

\multiput(30,0)(7.5,0){5}{\circle*{1}}

\put(-2,-15){\footnotesize{1}}
\put(28,-15){\footnotesize{2}}
\put(53,-15){\footnotesize{p--2}}
\put(83,-15){\footnotesize{p--1}}
\put(118,17){\footnotesize{p+1}}
\put(118,-25){\footnotesize{p}}

\end{picture}
\\

\hline
\end{tabular}
\end{center}
\label{Fig1}
\end{figure}
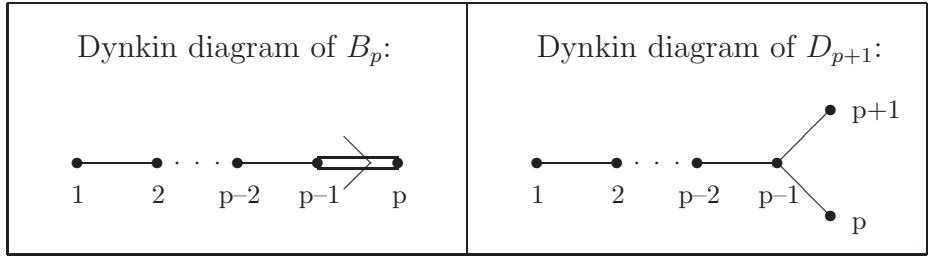

We now consider embeddings of the complex algebras (i.e. $B_p\subset D_{p+1}$), and later show that the embeddings hold for the real forms. Studying the Dynkin diagrams in Figure \ref{Fig1} suggests the construction.
\eqnb
    K^i & \equiv & H^i \phantom{\al_i\al_i\al_i\al_i\al_i} 1 \leq i\leq p-1\no
    F^{\pm\be^{(i)}}& \equiv & E^{\pm\al^{(i)}}\phantom{\al\al\al_i\al_i}1 \leq i\leq p-1\no
    F^{\pm\be^{(p)}}& \equiv & E^{\pm\al^{(p)}} + E^{\pm\al^{(p+1)}}\no
    K^p & \equiv & H^p + H^{p+1},
    \label{embedding3}
\eqne
where $K^i$ and $F^{\pm\be^{(i)}}$, $i=1,\ldots , p$, generate $B_p$. It is readily checked that eq.\ (\ref{embedding3}) generates $B_p$ by checking the relations
\eqnb
    [K^i, F^{\pm\be^{(j)}}] & = & \pm  ({A_{B_p}})^{ji} F^{\pm\be^{(j)}} \no
    [ F^{\be^{(i)}}, F^{-\be^{(j)}}] & = & \de^{ij}K^i\no
    \left({\rm ad}_{ F^{\pm\be^{(i)}}}\right)^{1-{ A}^{ji}} F^{\pm\be^{(j)}} & = & 0 \quad i\neq j ,
\eqne
where ${\rm A}^{ij}_{B_p}$ is the the Cartan matrix for $B_p$. Furthermore, by using the Hermitean conjugation rules with respect to $\mf{so}(2p,2)$, we also get the correct real form of $B_p$, $\mf{so}(2p,1)$. 

The non-compact roots w.r.t.\ the real form $\mf{so}(2p,1)$ lie in a $p$-dimensional space. Viewed as a subspace of the $p+1$-dimensional space, they correspond to vectors $\beta^{\{r\}} \equiv \frac{1}{2}\left(\al^{\{r\}}+\al^{\prime\{r\}}\right)$, as can be seen by using the orthonormal basis and eq.\ (\ref{D-non-compact}). The choice of basis of the roots of $B_p$ is such that there is a unique simple non-compact root given by $\beta^{(p)}$. This implies that the highest root $\theta_{B_{p}}$ is compact since the Coxeter label corresponding to this root is $2$. Furthermore, in this basis $|\theta_{B_{p}}|^2=2$, so that the Dynkin index of the embedding is one. We will denote by $\tilde\Delta$ the roots of $B_p$. Introduce the notation 
\eqnb
\left(\theta,\al\right)'\equiv \sum_{i=1}^{p-1}\theta_i\left(\alpha^{(i)\vee},\alpha\right)+\theta_p\left(\alpha^{(p)\vee}+\alpha^{(p+1)\vee},\alpha\right),
\eqne
where the scalar products on the righthand side are in the root space of $D_{p+1}$. By using the orthonormal basis above, one easily shows the following
\eqnb
\prod_{\al \in \Delta_c^+}\frac{1}{1-q^ne^{-i(\theta, \al)'}}=\prod_{\al \in \tilde{\Delta}_c^+}\frac{1}{1-q^ne^{-i(\theta, \al)}}
\label{eq27}
\eqne
\eqnb
\prod_{\al \in \Delta_n^+}\frac{1}{1-q^ne^{-i(\theta, \al)'}}=\prod_{\al \in \tilde{\Delta}_n^+}\left[\frac{1}{1-q^ne^{-i(\theta, \al)}}\right]^2
\label{eq28}
\eqne
where $q$ and $\theta_i$, $i=1,\ldots , p$, are arbitrary parameters. These particular relations will be used in the next section. 

Let us discuss the above construction for the coset ${SO}(2p-1,2)/{SO}(2p-1,1)$. $\mf{so}(2p-1,2)$ is a real form of $B_p$ and $\mf{so}(2p-1,1)$ is a real form of $D_p$. The construction above relies on using the complex algebra and projecting onto a real form by taking roots to be either compact or non-compact. This implies that the Cartan generators are compact, which in turn implies that the number of non-compact generators are even. This is not the case for $\mf{so}(2p-1,1)$, which has $2p-1$ non-compact generators. Therefore, one cannot apply this construction to the even-dimensional case.

In order to gauge the $\mathfrak{so}(2p,1)$ degrees of freedom, we use the formalism of Karabali and Schnitzer \cite{Karabali:1989dk}. This approach uses the BRST symmetry to define the coset space. As was shown in \cite{Bjornsson:2007ha}, the BRST formulation is necessary to achieve unitarity. In order to construct a nilpotent BRST charge one starts with the $\mathfrak{so}(2p,2)$ WZNW model at level $k$ and supplements it with an auxiliary sector, which is a $\mathfrak{so}(2p,1)$ WZNW model of level $\tilde k = -k-2g^{\vee}_{\mathfrak{so}(2p,1)}=-k-4p+2$. We will denote the corresponding current modes by $\tilde K_n^i$ and $ \tilde F^{\al}_n$, where $i=1,\ldots, p$ and $\al\in\Delta_c$. We define $\tilde{{\cal H}}_{\tilde{\mu}}$ to be the (irreducible) state-space over a highest weight state with weight $\hat{\tilde\mu}$. 

From the commutators of the subalgebra, see eq.\ (\ref{affalgebra}), one can determine a BRST-charge 
\eqnb
Q_1
      &=&
          \sum_{i=1}^{p}\sum_{n\in\Z}:c_{i,-n}\left(K_n^i + \tilde K_n^i\right): + \sum_{n\in\Z,\al\in\tilde\Delta}:c_{-n}^{\al}\left(F^{-\al}_n
          + \tilde F^{-\al}_n\right):
      \no 
      &+&
          \sum_{i=1}^{p}\sum_{\alpha\in\tilde\Delta}\sum_{m,n\in \Z}\al^i:c_{i,m} c^\al_n b^{-\al}_{-m-n}: 
      \no
      &-&
          \frac{1}{2}\sum_{\al,\be\in\tilde\Delta}\sum_{m,n\in \Z}\left[e_{\al,\be}:c^{-\al}_m c^{-\be}_n b^{\al+\be}_{-m-n}:
          + \,\delta_{\al+\be,0}\sum_{i=1}^{p}\al^{\vee}_i:c^{-\al}_m c^{\al}_n b^{i}_{-m-n}:\right],
\label{brst-coset}
\eqne
where $:\ldots :$ denotes normal ordering and we have introduced the $bc$-ghosts with the non-zero brackets
\eqnb
\left[c_{m,i},b_{n}^j\right]
      &=&
          \delta_{m+n,0}{\delta_i}^j
      \no
\left[c_{m}^\al,b_{n}^\be\right]
      &=&
          \delta_{m+n,0}\delta^{\al+\be,0}.      
\eqne
It is conventional to define the following ghost "vacuum"
\eqnb
b_m^i\left| 0\right>_{b,c}=b_p^\al\left| 0\right>_{b,c}
      &=&
          0\no
c_{n,i}\left| 0\right>_{b,c}=c_q^\al\left| 0\right>_{b,c}&=&
          0,
\label{ghostvac}
\eqne
for $m\geq 0$; $n>0$; $\al\in\tilde{\Delta}^+$, $p\geq 0$ and $q\geq 0$;  $\al\in\tilde{\Delta}^-$, $p>0$ and $q> 0$. The state-spaces spanned the $bc$-ghosts by acting with $bc$-creation operators is denoted by ${\cal H}^{bc}$. The Hermiticity properties are defined to be $(b^\al_n)^\dagger=\pm b^{-\al}_{-n}$, $(c^\al_n)^\dagger=\pm c^{-\al}_{-n}$, $(b^i_n)^\dagger=b^{i}_{-n}$ and $(c_{n,i})^\dagger=c_{-n,i}$, where the plus (minus) sign refers to compact (non-compact) roots $\al$.

The coset construction is now formulated through the BRST condition, so that states in the coset space satisfy
\eqnb
Q_1\left| S\right>
      &=&
          0
\no
b_0^i\left| S\right>
      &=&
          0.\phantom{1234}i=1,\ldots,p
\label{cosetcondition}
\eqne
States satisfying these equations and that are non-trivial in the $Q_1$ cohomology, i.e.\ non-exact, are true states in the coset model. Now eq.~(\ref{cosetcondition}) does not represent physical states in our case, since the string theory, that is represented by this WZNW model and possibly some unitary conformal field theory coupled to it, is defined by including the Virasoro conditions. Thus, we define the full BRST operator
\eqnb
Q
        &=&
            Q_1+\sum_{n\in\Z}\left(L_n+\tilde L_n+L_n'-\delta_{n,0}\right)\eta_{-n}
            - \sum_{m,n\in\Z}m:\eta_{-m}\eta_{-n}\mathcal{P}_{m+n}:
        \no
        &+&
            \sum_{m,n\in\Z}\sum_{\al\in\tilde{\Delta}}\left(n:\eta_{-m}c^{-\al}_{-n}b^{\al}_{m+n}:+n:\eta_{-m}c_{-n,i}b^{i}_{m+n}:\right).
\label{BRST charge}            
\eqne
$L'_n$ originates from some unitary CFT and $(\eta,\mathcal{P})$ are the usual conformal ghosts.

The $(\eta,{\mathcal{P}})$-ghost state-space ${\cal H}_{\eta{\mathcal{P}}}$ is defined as
for the $bc$-ghosts, with a "vacuum" state 
\eqnb
\mathcal{P}_{m}\left| 0\right>_{\eta,\mathcal{P}}
      &=&
          0\no
\eta_{n}\left| 0\right>_{\eta,\mathcal{P}}
      &=&
          0,
\label{ghostvac2}
\eqne
for $m\geq 0$ and $n>0$. The corresponding state-space is denoted by ${\cal H}^{\eta\mathcal P}$. The full ghost "vacuum" is the product of the two separate ghost parts, $\left| 0\right>_{ghost}=\left|0\right>_{b,c}\otimes\left|0\right>_{\eta,\mathcal{P}}$. We denote the product space by ${\cal H}^{ghost}={\cal H}^{bc}\times{\cal H}^{\eta\mathcal P}$. We denote by ${\cal H'}^{ghost}$ the subspace of states satisfying $b^i_0\left| \Phi\right>=0$, $i=1,\ldots,p$, and $\mathcal{P}_0\left| \Phi\right>=0$.

Using the BRST operators one can construct the following BRST exact quantities
\eqnb
K_0^{{\mathrm{tot}},i}
      &\equiv&
           \left[Q,b^i_0\right]=
          K^i_0+\tilde K^i_0+\sum_{\al,m}\al^i:b^\al_{-m}c^{-\al}_m:
\label{H tot}
      \\
L^{\mathrm{tot}}_n
&\equiv&
           \left[Q,\mathcal{P}_n\right]=
           L_n+\tilde L_n+L'_n+L^{gh}_n-\delta_{n,0},\label{exact-curr}
\label{L tot}
\eqne
where
\eqnb
L_n
      &=&
          \frac{1}{2\left(k+2p\right)}\sum_{m\in\Z}\left(:{G}_{ij}H^i_{m}H^j_{n-m}:
          + \sum_{\al\in\Delta}\frac{\left(\al,\al\right)}{2}:E^{-\al}_{m}E^{\al}_{n-m}:\right)
      \no
\tilde L_n
      &=&
          -\frac{1}{2\left(k+2p-1\right)}\sum_{m\in\Z}\left(:\tilde {G}_{ij}\tilde K^i_{m}\tilde K^j_{n-m}:
          + \sum_{\al\in\tilde\Delta}\frac{\left(\al,\al\right)}{2}:\tilde F^{-\al}_{m}\tilde F^{\al}_{n-m}:\right)
      \no
L^{gh}_n
      &=&
          \sum_{m\in\Z}\left(m:b^i_{n-m}c_{m,i}:+\sum_{\al\in\tilde\Delta} m:b^{-\al}_{n-m}c^\al_{m}:\right).
\eqne

Using $Q$ as our BRST charge, the physical state-space of the string theory is defined by the conditions
\eqnb
Q\left| \Phi\right>
      &=&
          0
\no
b^i_0\left| \Phi\right>
      &=&
          0\phantom{1234}i=1,\ldots, p
\no
\mathcal{P}_0\left| \Phi\right>
      &=&
          0.
\label{brstcondition}
\eqne
We denote by ${\cal H}^{Q}_{\hat{\mu}\tilde{\hat{\mu}}}$ the sub-space of states of $\mathcal{H}_{\hat\mu} \times\tilde{\mathcal{H}}_{\hat{\tilde{\mu}}}\times\mathcal{H}^{\mathrm{CFT}}_{l'} \times\mathcal{H}^{ghost}$ satisfying the above equations. Here $\mathcal{H}^{\mathrm{CFT}}_{l'}$ represents some unitary CFT. States that are in ${\cal H}^{Q}_{\hat{\mu}\tilde{\hat{\mu}}}$ will be called trivial if they are $Q$-exact and non-trivial if they are not.

Non-trivial states in ${\cal H}^{Q}_{\hat{\mu}\tilde{\hat{\mu}}}$ have to satisfy 
\eqnb
K_0^{{\mathrm{tot}},i}\left|\Phi\right>&=&0,\phantom{1234}i=1,\ldots, p \label{relcond}\\ 
L_0^{{\mathrm{tot}}}\left|\Phi\right>&=&0,\label{L0}
\eqne
which follows directly from eq.~(\ref{brstcondition}) by taking the commutator of $Q$ with $b_0^i$ and $\mathcal{P}$, respectively. 

The representations that we will focus on in this work are antidominant highest weight representations for $D_{p+1}$. We believe that these are relevant for the string theories that we consider here as was well as in the earlier treated models in \cite{Bjornsson:2007ha}--\cite{Bjornsson:2009dp}.

For the auxiliary sector, the class of representations that are natural are found by studying the requirement that there should exist conventional BRST invariant ground-states. Such states are of the form
\eqnb
\left| 0;\mu,\tilde \mu\right> 
      &\equiv&
          \left| 0;\mu\right> \otimes \left| \tilde 0;\tilde \mu\right> \otimes \left| 0\right>_{ghost}.
\eqne
Using eq.~(\ref{relcond}) we have
\eqnb
0=K_0^{{\mathrm{tot}},i}\left| 0;\mu,\tilde \mu\right>
      &=&
          (\mu^i+\tilde {\mu}^i+2)\left| 0;\mu,\tilde \mu \right>, \ \ i=1,\ldots, p-1,\label{eq H0}
\eqne
and
\eqnb
0=K_0^{{\mathrm{tot}},p}\left| 0;\mu,\tilde \mu\right>
      &=&
          (\mu^p+\mu^{p+1}+\tilde {\mu}^p+2)\left| 0;\mu,\tilde \mu \right>.\label{eq H1}
\eqne
This implies that if we choose $\mu$ to be antidominant we must require $\tilde{\mu}$ to be a dominant $B_p$ weight. This implies that the auxiliary sector has representations that are not unitary. Note that one may straightforwardly show that one needs to have an antidominant component of $\mu$ in the non-compact direction.

For a general non-trivial state  in ${\cal H}^{Q}_{\hat{\mu}\tilde{\hat{\mu}}}$ we have to require eq.~(\ref{relcond}), which implies the following. 
\begin{lemma} ${\cal H}^{Q}_{\hat{\mu}\hat{\tilde{\mu}}}$ is trivial unless
\eqnb
&~&\mu^i+\tilde {\mu}^i+2\tilde \rho^i-\sum_{j=1}^{p+1}m_j\al^{(j)i}-\sum_{j=1}^{p}\tilde{m}_j\be^{(j)i}=0,
\hs{3mm}i=1,\ldots, p-1\no
&~&\mu^p+\mu^{p+1}+\tilde {\mu}^p+2\tilde \rho^p-\sum_{j=1}^{p+1}m_j\left(\al^{(j)p}+\al^{(j)p+1}\right)-
\sum_{j=1}^{p}\tilde{m}_j\be^{(j)i}=0
\eqne
for some integers $m_j$ and $\tilde{m}_j$. Here $\al^{(j)}\in\Delta_s$, $\be^{(j)}\in\tilde{\Delta}_s$. In particular, unless $(\mu^i+\tilde {\mu}^i)\in\mathbb{Z}$, $i=1,\ldots,p-1$, and $\mu^p+\mu^{p+1}+\tilde {\mu}^p\in\mathbb{Z}$, ${\cal H}^{Q}_{\hat{\mu}\hat{\tilde{\mu}}}$ is trivial. \label{lemma 1}
\end{lemma}


\sect{Unitarity}
In order to analyze unitarity of the space of states satisfying eq.~(\ref{brstcondition}), we use the same technique that we successfully applied previously in \cite{Bjornsson:2007ha}. Define, therefore, the character
\eqnb
\chi^{\mathrm{tot}}(\tau,\theta)
      &\equiv& 
          \Tr\left[\exp\left[2\pi\ii\tau\left(L^{\mathrm{tot}}_0\right)\right]
          \exp\left[\ii\sum_{i=1}^{p}\theta_i K_0^{{\mathrm{tot}},i}) \right](-1)^{\Delta N_{gh}}\right],
\label{defchar}
\eqne
$\Delta N_{gh}$ is the ghost number of the state in question relative to the ghost vacuum. The trace is taken over all states. The character decomposes into separate parts
\eqnb
\chi^{\mathrm{tot}}\left(\tau,\theta\right)
      &=&
          e^{-2\pi\ii \tau}\chi^{\hat{\mf{so}}(2p,2)}\left(\tau,\theta\right)\tilde\chi^{\hat{\mf{so}}(2p,1)}
          \left(\tau,\theta\right)\chi^{\mathrm{CFT}}(\tau)\chi^{gh}\left(\tau,\theta\right)\chi^{CFT\;gh}\left(\tau\right).
          \label{char-decomposed}
\eqne
The character defined in eq.~(\ref{defchar}) receives only contributions from non-trivial BRST invariant states. However, since the physical states not only satisfy the BRST condition, but all the conditions in eqs.~(\ref{brstcondition}), (\ref{relcond}) and (\ref{L0}), we must instead consider the following function
\eqnb
\int d\tau\mathcal  B^{\left(\hat{\mf{so}}(2p,2),\,\hat{\mf{so}}(2p,1)\oplus Vir\right)}(\tau)
      &\equiv&
          \int d\tau\int d\theta\hspace{1mm}\left\{ \chi^{\mathrm{tot}}(\tau,\theta)\right\}
      \no
      & \hspace*{-95mm}\equiv&
          \hspace*{-49mm}
          \int d\tau\int d\theta \hspace{1mm}\Tr\left[\exp\left[2\pi\ii\tau\left(L^{\mathrm{tot}}_0\right)\right]
          \exp\left[\ii\sum_{i=1}^{p}\theta_i K_0^{{\mathrm{tot}},i}\right](-1)^{\Delta N_{gh}}\right]\!\!,
\eqne
where we have defined $d\theta = \prod_{i=1}^p d\theta_i$. The $\tau$- and $\theta$-integrations are formal integrations which act as projections, $\int d\tau \int d\theta e^{2\pi \ii \tau p}e^{i\theta r}=\delta_{p,0}\delta_{r,0}$, which is required by the eqs.~(\ref{relcond}) and (\ref{L0}). The function $ B^{\left(\hat{\mf{so}}(2p,2),\,\hat{\mf{so}}(2p,1)\oplus Vir\right)}(\tau,\phi)$ was first introduced in \cite{Hwang:1993nc} and extended in \cite{Hwang:1994yr}. We also define another function, the signature function,
\eqnb
\Sigma^{\mathrm{tot}}(\tau,\theta) 
	&\equiv&
         \Tr'\left[\exp\left[2\pi\ii\tau\left(L^{\mathrm{tot}}_0\right)\right]
         \exp\left[\ii\sum_{i=1}^{p}\theta_i K_0^{{\mathrm{tot}},i}\right](-1)^{\Delta N_{gh}}\right].
\label{defsignature}
\eqne
The prime on the trace indicates that the trace is taken with signs i.e.\ a state with positive (negative) norm contributes with a positive (negative) sign in the trace. We define a corresponding coset signature function
\eqnb
\int d\tau\mathcal S^{\left(\hat{\mf{so}}(2p,2),\,\hat{\mf{so}}(2p,1)\oplus Vir\right)}(\tau)
      &\equiv&
          \int d\tau\int d\theta\hspace{1mm}\left\{ \Sigma^{\mathrm{tot}}(\tau,\theta)\right\}
      \no
      & \hspace*{-90mm}\equiv&
          \hspace*{-45mm}
          \int d\tau\int d\theta\hspace{1mm}\Tr'\left[\exp\left[2\pi\ii\tau\left(L^{\mathrm{tot}}_0\right)\right]
          \exp\left[\ii\sum_{i=1}^{p}\theta_i K_0^{{\mathrm{tot}},i}\right](-1)^{\Delta N_{gh}}\right]\!\!.
      \label{cosetsignature}
\eqne
Since the projection of the character onto states satisfying eqs.~(\ref{relcond}) and (\ref{L0}) gives the total number of states in ${\cal H}^{Q}_{\hat{\mu}\tilde{\hat{\mu}}}$ for given weights and the same projection of the signature function gives the difference between the number of positive and negative norm states in the same state-space, we have the following lemma. 
\begin{lemma} 
${\cal H}^{Q}_{\hat{\mu}\tilde{\hat{\mu}}}$ is unitary if, and only if, 
\eqnb
\int d\tau\left[\mathcal{B}^{\left(\hat{\mf{so}}(2p,2),\,\hat{\mf{so}}(2p,1)\oplus Vir\right)}(\tau,\phi)-
\mathcal{S}^{\left(\hat{\mf{so}}(2p,2),\,\hat{\mf{so}}(2p,1)\oplus Vir\right)}(\tau,\phi)\right]=0.
\label{eq328}
\eqne
\label{lemma 2}
\end{lemma}
The explicit form of the characters and signature functions involved are given by the following lemmas where we have defined $q\equiv e^{2\pi\ii \tau}$.
\begin{lemma} 
Let $\hat\mu$ be an antidominant weight. The character for the combined $\hat{\mf{so}}(2p,2)$- and ghost-sectors is
\eqnb
\chi^1_\mu(\tau,\theta) 
      &=&
          q^{\frac{\mathcal{C}_2(\mu)}{2\left(k+2p\right)}} 
          e^{\ii(\theta,\mu)'+\ii\left(\theta,2\tilde{\rho}\right)}
          \prod_{\alpha\in\tilde\Delta^+_c}\left(1-e^{-\ii\left(\theta,\alpha\right)}\right)
          \prod_{m=1}^{\infty}\left(1-q^m\right)^{p+1}\prod_{\alpha\in\tilde{\Delta}_c}\left(1-q^m e^{\ii\left(\theta,\alpha\right)}\right)
        .\no
\label{char1}
\eqne
The corresponding signature function is 
\eqnb
\Sigma^1_\mu(\tau,\theta) 
       &=&
                q^{\frac{\mathcal{C}_2(\mu)}{2\left(k+2p\right)}} 
                e^{\ii(\theta,\mu)'+\ii\left(\theta,2\tilde\rho\right)}
                \prod_{\alpha\in\tilde\Delta^+_c}\left(1-e^{-\ii\left(\theta,\alpha\right)}\right)
                \prod_{\alpha\in\tilde{\Delta}^+_n}\frac{1+e^{-\ii\left(\theta,\alpha\right)}}{1-e^{-\ii\left(\theta,\alpha\right)}}
            \no
            &\times&
                \prod_{m=1}^{\infty}\left\{
                   \left(1-q^m\right)^{p+1} \prod_{\alpha\in\tilde\Delta_c}(1-q^m e^{\ii\left(\theta,\alpha\right)}) \prod_{\alpha\in\tilde{\Delta}_n}\frac{1+q^m e^{\ii\left(\theta,\alpha\right)}}{1-q^m e^{\ii\left(\theta,\alpha\right)}}\right\},
\label{sign1}
\eqne
where $\mathcal{C}_{2}\left(\mu\right)$ is the quadratic Casimir of $D_{p+1}$.
\label{lemma 3}
\end{lemma}
{\bf Proof.} The characters and signature functions for the different sectors were essentially derived in \cite{Bjornsson:2007ha}. The difference compared to here is the choice of parameters describing the functions. In the present case we have 
\eqnb
\chi^{\mf{so}(2p,2)}_\mu(\tau,\theta) 
      &=&
          q^{\frac{\mathcal{C}_2(\mu)}{2\left(k+2p\right)}} 
          e^{\ii(\theta,\mu)}
          \prod_{\alpha\in\tilde{\Delta}^+_c}\frac{1}{1-e^{-\ii\left(\theta,\alpha\right)}}
          \prod_{\alpha\in\tilde{\Delta}^+_n}\frac{1}{\left(1-
          e^{-\ii\left(\theta,\alpha\right)}\right)^2}
      \no
      &\hspace{-5mm}\times&
	\hspace{-5mm}
          \prod_{m=1}^{\infty}
          \left\{
            \frac{1}{ \left(1-q^m\right)^{p+1}}
             \prod_{\alpha\in\tilde{\Delta}_c} \frac{1}{ 1-q^m e^{\ii\left(\theta,\alpha\right)}}
          \prod_{\alpha\in\tilde{\Delta}_n}\frac{1}{\left(1-q^m
          e^{\ii\left(\theta,\alpha\right)}\right)^2}
          \right\},
\label{char-g}
\eqne
\eqnb
\Sigma^{\mf{so}(2p,2)}_\mu(\tau,\theta) 
      &=&
          q^{\frac{\mathcal{C}_2(\mu)}{2\left(k+2p\right)}} 
          e^{\ii(\theta,\mu)}
          \prod_{\alpha\in\tilde{\Delta}^+_c}\frac{1}{1+e^{-\ii\left(\theta,\alpha\right)}}
          \prod_{\alpha\in\tilde{\Delta}^+_n}\frac{1}{\left(1-
          e^{-\ii\left(\theta,\alpha\right)}\right)^2}
      \no
      &\hspace{-5mm}\times&
	\hspace{-5mm}
          \prod_{m=1}^{\infty}
          \left\{
            \frac{1}{ \left(1+q^m\right)^{p+1}}
             \prod_{\alpha\in\tilde{\Delta}_c} \frac{1}{ 1+q^m e^{\ii\left(\theta,\alpha\right)}}
          \prod_{\alpha\in\tilde{\Delta}_n}\frac{1}{\left(1-q^m
          e^{\ii\left(\theta,\alpha\right)}\right)^2}
          \right\},
\label{sign-g}
\eqne
where we have used eqs.\ (\ref{eq27}) and (\ref{eq28}). For the combined ghost sectors we have
\eqnb
\chi^{gh}\left(q,\theta\right)
      &=&
          e^{\ii\left(\theta,2\tilde\rho\right)}
	  \prod_{\alpha\in\tilde{\Delta}^+}
	  \left(1-e^{-\ii\left(\theta,\alpha\right)}\right)^2
      \no
      &\times&
          \prod_{m=1}^\infty \left\{
              \left(1-q^m\right)^{2p+2}
              \prod_{\alpha\in\tilde\Delta}
	      \left(1-q^m e^{\ii\left(\theta,\alpha\right)}\right)^2\right\},
\label{char-ghost}
\\
\Sigma^{gh}\left(q,\theta\right)
      &=&
          e^{\ii\left(\theta,2\tilde\rho\right)}
	  \prod_{\alpha\in\tilde{\Delta}^+}
	  \left(1-e^{-\ii\left(\theta,\alpha\right)}\right)\left(1+e^{-\ii\left(\theta,\alpha\right)}\right)
      \no
      &\times&
          \prod_{m=1}^\infty \left\{
              \left(1-q^m\right)^{p+1}\left(1+q^m\right)^{p+1}
              \prod_{\alpha\in\tilde\Delta}
	      \left(1-q^m e^{\ii\left(\theta,\alpha\right)}\right)\left(1+q^m e^{\ii\left(\theta,\alpha\right)}\right)\right\},
\no
\label{sign-ghost}
\eqne
The lemma now follows by multiplying the above functions together. $\Box$

We will also need expressions for the character and the signature function corresponding to $\mf{so}(2p,1)$. 
\begin{lemma}
Let $\hat{\tilde{\mu}}$ be a dominant integral weight. The character for an irreducible affine $B_p$-module of highest weight
$\hat{\tilde{\mu}}$ is given by 
\eqnb
\chi^{\mf{so}(2p,1)}_{\tilde\mu}(\tau,\theta) 
      &=&
          q^{-\frac{\tilde{\mathcal{C}}_2(\tilde\mu)}{2\left(k+2p-1\right)}} 
          e^{\ii(\theta,\tilde\mu)}
          \prod_{\alpha\in\tilde\Delta^+_c}\frac{1}{1-e^{-\ii\left(\theta,\alpha\right)}}
          \prod_{\alpha\in\tilde\Delta^+_n}\frac{1}{1-
          e^{-\ii\left(\theta,\alpha\right)}}
      \no
      &\times&
          \prod_{m=1}^{\infty}\left\{
            \frac{1}{ \left(1-q^m\right)^{p}}\prod_{\alpha\in\tilde\Delta_c} \frac{1}{ 1-q^m e^{\ii\left(\theta,\alpha\right)}}  
          \prod_{\alpha\in\tilde\Delta_n}\frac{1}{1-q^m
          e^{\ii\left(\theta,\alpha\right)}}
          \right\}
          \no
          &\times& 
          \sum_{w\in \tilde W} \sum_{\be\in \tilde{L}^\vee}
          {\sign(w)}
          e^{\ii \left(w\left(\tilde\mu+\tilde\rho-\be\left(k+2p-1\right)\right)-\tilde\mu-\tilde\rho,\theta\right)}
      \no
      &\times&
          q^{\left(\be,\tilde\mu+\tilde\rho\right)-\frac{1}{2}\left(\be,\be\right)\left(k+2p-1\right)}.
          \label{char-h}
\eqne
The signature function w.r.t.\ the real form $\hat{\mf{so}}(2p,1)_{\tilde k}$ is
\eqnb
\Sigma^{\mf{so}(2p,1)}_{\tilde\mu}(\tau,\theta) 
          &=&
          q^{-\frac{\tilde{\mathcal{C}}_2(\tilde\mu)}{2\left(k+2p-1\right)}} 
          e^{\ii(\theta,\tilde\mu)}
          \prod_{\alpha\in\tilde\Delta^+_c}\frac{1}{1-e^{-\ii\left(\theta,\alpha\right)}}
          \prod_{\alpha\in\tilde\Delta^+_n}\frac{1}{1-
          e^{-\ii\left(\theta,\alpha\right)}}
      \no
      &\times&
          \prod_{m=1}^{\infty}\left\{
            \frac{1}{ \left(1-q^m\right)^{p}}\prod_{\alpha\in\tilde\Delta_c} \frac{1}{ 1-q^m e^{\ii\left(\theta,\alpha\right)}}  
          \prod_{\alpha\in\tilde\Delta_n}\frac{1}{1-q^m
          e^{\ii\left(\theta,\alpha\right)}}
          \right\}
          \no
          &\times&
          \sum_{w\in \tilde W} \sum_{\be\in \tilde{L}^\vee}
          {\sign(w)}(-1)^{2\left(w\left(\tilde\mu+\tilde\rho\right)-\tilde\mu-\tilde\rho,
                 \tilde{\Lambda}_{(p)}\right)}
                 \no
          &\times& 
          e^{\ii \left(w\left(\tilde\mu+\tilde\rho-\be\left(k+2p-1\right)\right)-\tilde\mu-\tilde\rho,\theta\right)}
          q^{\left(\be,\tilde\mu+\tilde\rho\right)-\frac{1}{2}\left(\be,\be\right)\left(k+2p-1\right)},
\label{sign-h}
\eqne
where $\tilde{\mathcal{C}}_{2}\left(\tilde{\mu}\right)$ is the quadratic Casimir of $B_{p}$.
\label{lemma5}
\end{lemma}
{\bf Proof.} The character above follows directly from the Weyl-Kac character formula \cite{Kac:1974}. In order to prove the expression for the signature function, we proceed in several steps. First, consider the following state of weight $\hat{\tilde{\mu}}'$
\eqnb
\left|\hat{\tilde\mu}'\right>\equiv \left[F^{-\hat\be^{(0)}}\right]^{m_0}\left[F^{-\hat\be^{(1)}}\right]^{m_1}
\ldots \left[F^{-\hat\be^{(p)}}\right]^{m_{p}}
\left|\hat{\tilde\mu}\right>.
\label{eq337}
\eqne
where we, for simplicity, have used affine notation, with $\hat\be^{(0)}$ being the additional affine simple root. This state, if it is in the irreducible module, has a signature $(-1)^{m_p}$. This follows since $F^{\hat\be^{(0)}}$, $F^{\hat\be^{(1)}}$, ... , $F^{\hat\be^{(p-1)}}$ and $\ii F^{\hat\be^{(p)}}$  satisfy Hermiticity conditions corresponding to $\hat{\mf{so}}(2p+1)_{\tilde k}$ and $\hat{\tilde\mu}$ being dominant implies unitary highest weight representations for $\hat{\mf{so}}(2p+1)_{\tilde k}$. One has
\eqnb
m_{p}
&=&
2\left( \hat\Lambda_{(p)}', \hat{\tilde\mu}'-\hat{\tilde\mu}\right) \phantom{12}\mathrm{mod}\;2
\no
&=&
2\left(\tilde\Lambda_{(p)}, {\tilde\mu}'-{\tilde\mu}\right) \phantom{12}\mathrm{mod}\;2,
\label{eq338}
\eqne
with $\hat\Lambda_{(p)}'$ defined as the weight $(\tilde\Lambda_{(p)}, 0, 0)$. This follows as $\be^{(i)\vee}=2\be^{(i)}$, $(\tilde\Lambda_{(p)},\be^{(i)\vee})=\delta^i_p$ and $(\tilde\Lambda_{(p)},\tilde\theta)\in\Z$ with $\tilde\theta$ being the highest root of $B_p$.

Now, a general state of a definite weight $\hat{\tilde\mu}'$ is given by a linear combination of states of the form eq.~(\ref{eq337}), but with an arbitrary ordering among the generators. We may again apply the argument that such a state has positive norm if we replace the non-compact generators with $i$ times the non-compact generator. Hence, also in this case we arrive at the signature $(-1)^{m_{p}}$, with $m_{p}$ given by eq.~(\ref{eq338}).

Turning now to the signature function, the above result implies that the signature differs from the character by a sign factor $(-1)^{2\left(\tilde\Lambda_{(p)}, {\tilde\mu}'-{\tilde{\mu}}\right)}$ for a term in the character corresponding to a state of weight ${\tilde\mu}'$. The weight can easily be read off from the character and, in particular, it is straightforward to read off the $p$'th  component. The correct sign factor comes from taking $\theta\rightarrow \theta+2\pi\tilde\Lambda_{(p)}$ together with a compensating sign factor for the highest weight state.   Thus, a factor $\left(1-q^n\exp(-\ii(\theta,\be))\right)^{-1}$ turns into $\left(1+q^n\exp(-\ii(\theta,\be))\right)^{-1}$ for non-compact roots, whereas the factor remains unchanged for compact roots. Similarly, terms of the form 
\eqnb
{\sign( w)}e^{\left(\theta,w(\tilde\mu+\tilde\rho+\tilde\beta(k+2p-1))-\tilde\mu-\tilde\rho\right)}
\eqne
will change into 
\eqnb
{\sign( w)}(-1)^{2\left(w(\tilde\mu+\tilde\rho+\tilde\beta(k+2p-1))-\tilde\mu-\tilde\rho,\tilde{\Lambda}_{(p)}\right)}e^{\left(\theta,w(\tilde\mu+\tilde\rho+\tilde\beta(k+2p-1))-\tilde\mu-\tilde\rho\right)}.
\eqne
This can be simplified by using that the coroot lattice is invariant under Weyl group and that $\left(\beta,\tilde{\Lambda}_{(p)}\right) \in \Z$. Therefore, the previous equation can be simplified as
\eqnb
{\sign( w)}(-1)^{2\left(w(\tilde\mu+\tilde{\rho})-\tilde\mu-\tilde\rho,\tilde{\Lambda}_{(p)}\right)}e^{\theta,w(\tilde\mu+\tilde\rho+\tilde\beta(k+2p-1))-\tilde\mu-\tilde\rho}.
\eqne 
This proves the lemma. \ $\Box$

We now state the main result of this paper.
\begin{theorem}
Let $\hat{\mu}$ be an antidominant weight. If $\hat{\tilde\mu}$ is a dominant integral with $\tilde\mu^p$ odd, then ${\cal H}^{Q}_{\hat{\mu}\tilde{\hat{\mu}}}$ is unitary. 
\label{Theorem1}
\end{theorem}

\vspace{5mm}
\noindent
{\bf Proof.} We will prove that the stated conditions are sufficient for unitarity by making use of Lemma \ref{lemma 2}. We will show that under conditions assumed, the integrands in eq.~(\ref{eq328}) are equal. If $\mu^i$, $i=1,\ldots, p-1$, or $\mu^{p}+\mu^{p+1}$, is not an integer then the state-space ${\cal H}^{Q}_{\hat{\mu}\tilde{\hat{\mu}}}$ is trivial by Lemma \ref{lemma 1}. Combining the characters and signature functions in eqs.~(\ref{char1}), (\ref{sign1}), (\ref{char-h}) and (\ref{sign-h}) we get
\eqnb
\chi^{\mathrm{tot}}_{\mu,\tilde\mu}(\tau,\theta) 
      &=&
          q^{\frac{\mathcal{C}_2(\mu)}{2\left(k+2p\right)}-\frac{\tilde{\mathcal{C}}_2(\tilde\mu)}{2\left(k+2p-1\right)}} 
          e^{\ii\left(\theta,\mu+\tilde\mu+2\tilde\rho\right)}
          \prod_{\alpha\in\tilde{\Delta}^+_n}\frac{1}{1-e^{-\ii\left(\theta,\alpha\right)}}
	\no
      &\times& 
            \prod_{m=1}^{\infty}\left\{\left(1-q^m\right)
            \prod_{\alpha\in\tilde{\Delta}_n}\frac{1}{1-q^m e^{\ii\left(\theta,\alpha\right)}}\right\}
	\no
       &\times&
                 \sum_{w\in \tilde W} \sum_{\be\in \tilde{L}^\vee}
                 {\sign(w)}
                 e^{\ii \left(w\left(\tilde\mu+\tilde\rho-\be\left(k+2p-1\right)\right)-\tilde\mu-\tilde\rho,\theta\right)}
             \no
             &\times&
          q^{\left(\be,\tilde\mu+\tilde\rho\right)-\frac{1}{2}\left(\be,\be\right)\left(k+2p-1\right)}
          \label{chartot}
 \eqne
\eqnb
\Sigma^{\mathrm{tot}}_{\mu,\tilde\mu}(\tau,\phi,\theta) 
      &=&
          q^{\frac{\mathcal{C}_2(\mu)}{2\left(k+2p\right)}-\frac{\tilde{\mathcal{C}}_2(\tilde\mu)}{2\left(k+2p-1\right)}} 
          e^{\ii\left(\theta,\mu+\tilde\mu+2\tilde\rho\right)}
          \prod_{\alpha\in\tilde{\Delta}^+_n}\frac{1}{1-e^{-\ii\left(\theta,\alpha\right)}}
	\no
      &\times& 
            \prod_{m=1}^{\infty}\left\{\left(1-q^m\right)
            \prod_{\alpha\in\tilde{\Delta}_n}\frac{1}{1-q^m e^{\ii\left(\theta,\alpha\right)}}\right\}
	\no
       &\times&
                 \sum_{w\in \tilde W} \sum_{\be\in \tilde{L}^\vee}
                 {\sign(w)}(-1)^{2\left(w\left(\tilde{\mu}+\tilde{\rho}\right)-\tilde{\mu}-\tilde{\rho},\tilde{\Lambda}_{(p)}\right)}
       \no
	&\times&
		e^{\ii \left(w\left(\tilde\mu+\tilde\rho-\be\left(k+2p-1\right)\right)-\tilde\mu-\tilde\rho,\theta\right)}
          q^{\left(\be,\tilde\mu+\tilde\rho\right)-\frac{1}{2}\left(\be,\be\right)\left(k+2p-1\right)}.
\eqne
Studying the above expressions, we see that they are equal provided the sign factor in the signature function satisfies,
\eqnb
(-1)^{2\left(w\left(\tilde{\mu}+\tilde{\rho}\right)-\tilde{\mu}-\tilde{\rho},\tilde{\Lambda}_{(p)}\right)} &=& 1.
\label{eq343}
\eqne
Let us, therefore, study under what circumstances this is the case. We will prove the following result. 
\begin{lemma}
Equation (\ref{eq343}) is satisfied if and only if $(\tilde{\mu},\be^{(p)\vee})$ 
is an odd integer.
\label{lemma6}
\end{lemma}
{\bf Proof.} For a fundamental Weyl reflection ${w}_{(i)}$ 
\eqnb
\tilde{\lambda}^{(i)}
	&\equiv& 
	{w}_{(i)}(\tilde{\mu}+\tilde\rho)-\tilde{\mu}-\tilde\rho
	\no
	&=&
	\left(\tilde{\mu}^i+1\right)\be^{(i)},
\eqne
we have 
\eqnb
2\left(\tilde{\lambda}^{(i)},\tilde{\Lambda}_{(p)}\right)
&=&\left(2-\delta_{p,i}\right)\left(\tilde{\mu}^i+1\right)\left(\be^{(i)\vee},\tilde{\Lambda}_{(p)}\right)
\no
&=&
\left\{\begin{array}{ll}
0               & \mathrm{for}\hs{3mm} i=1,\ldots, p-1\\
\tilde{\mu}^p+1 & \mathrm{for}\hs{3mm} i=p\end{array}\right..
\eqne
Consider first the 'only if' part of the lemma. If $\tilde{\mu}^p\notin2\Z+1$ then one can choose the fundamental Weyl reflection $w_{(p)}$, which yields that eq.~(\ref{eq343}) is not satisfied. We now study the 'if' part of the lemma. As fundamental Weyl reflections maps $\tilde{\mu}\stackrel{{w}_{(i)}}{\longrightarrow}\tilde{\mu} + n_i\beta^{(i)}$ (no summation over $i$), where $n_i\in\Z$, it is sufficient to prove that $n_i\beta^{(i)}$ satisfies $2n_i\left(w(\beta^{(i)}),\Lambda_{(p)}\right) \in 2\Z$ for all Weyl reflections $w$. As $n_p\in 2\Z$, $n_i\beta^{(i)}$ is an element in the coroot lattice. Using that the Weyl group is an automorphism of the coroot lattice and that $\left(\beta^{\vee},\Lambda_{(p)}\right)\in\Z$, we have proven the lemma.\ $\Box$

To conclude the proof of the theorem, dominant integral highest weights $\hat{\tilde{\mu}}$ imply that ${\tilde{\mu}}$ is integral. By Lemma \ref{lemma 1}, $\mu^i\in \Z$, $i=1,\ldots , p-1$ and $\mu^p+\mu^{p+1}\in \Z$. Observe also that the conditions of the theorem imply that $k$ is an integer. $\Box$

To generalize this to the world-sheet supersymmetric case is straightforward. This follows since this case is equivalent, by a field redefinition, to the bosonic model above and a free fermion model. The only difficulties come from the zero modes of the fermions in the Ramond-sector. This can be analyzed by the methods in \cite{Bjornsson:2008fq}. If one assumes that the highest weight is antidominant and that the representations of the auxiliary sector are integer dominant highest weight representations this yields, using the steps above and the methods in \cite{Bjornsson:2008fq}, that the model is unitary.


\sect{Discussion of the necessary conditions}
In this section we will discuss the problems to formulate and, thus, proving necessary conditions in this case. We will illustrate the difficulty of the problem by considering the component of the highest weight which is simplest to investigate, $\tilde{\mu}^{p}$, and restrict the analysis to the grade zero part of the algebra. Let us, furthermore, restrict our analysis by assuming that the highest weight for the auxiliary sector satisfies
\eqnb
\tilde{\mu}^{i} &>& -1 \no
\tilde{k} + 1 &>& \left(\tilde{\mu},\tilde\theta\right).
\label{eq41}
\eqne
One can expand the grade zero part of the character as
\eqnb
\chi_{\tilde{\mu}}^{\mf{so}(2p,1)}\left(\theta\right)
	&=& 
		e^{\ii\left(\theta,\tilde{\mu}\right)}\left[ \prod_{\alpha\in\tilde{\Delta}_+}\frac{1-\delta_{\tilde{\mu}^{p},\left[\tilde{\mu}^{p}\right]}e^{-\ii\left(\left[\tilde\mu^{p}\right]+1\right)\left(\theta,\beta^{(p)}\right)}}{1-e^{-\ii\left(\theta,\alpha\right)}}\right] + \mathcal{O}\left(e^{\ii\left(\gamma,\theta\right)}\right),
\label{eqn42}
\eqne
where we have excluded a factor proportional to $q^{(\ldots)}$, $[\cdot]$ denotes the integer part of and $\gamma$ is defined by $\left(\tilde{\mu}-\gamma,\Lambda_{(i)}\right) > 0$ for $1 \leq i \leq p-1$. The signature function is
\eqnb
\Sigma_{\tilde{\mu}}^{\mf{so}(2p,1)}\left(\theta\right)
	&=& 
		e^{\ii\left(\theta,\tilde{\mu}\right)}\left[ \prod_{\alpha\in\tilde{\Delta}^{c}_+}\frac{1}{1-e^{-\ii\left(\theta,\alpha\right)}}\prod_{\alpha\in\tilde{\Delta}^{n}_+}\frac{1}{1+e^{-\ii\left(\theta,\alpha\right)}}
	\right.
	\no
	&-&
		(-1)^{\left[\tilde{\mu}^{p}\right]+1}e^{-\ii\left(\left[\tilde\mu^{p}\right]+1\right)\left(\theta,\beta^{(p)}\right)}\prod_{\alpha\in\tilde{\Delta}^{c}_+}\frac{1}{1-e^{-\ii\left(\theta,\alpha\right)}}\prod_{\alpha\in\tilde{\Delta}^{n}_+}\frac{1}{1+e^{-\ii\left(\theta,\alpha\right)}}
	\no
	&+&
		\left(1-\delta_{\tilde{\mu}^p,\left[\tilde{\mu}^p\right]}\right)(-1)^{\left[\tilde{\mu}^{p}\right]+1}e^{-\ii\left(\left[\tilde\mu^{p}\right]+1\right)\left(\theta,\beta^{(p)}\right)}\frac{1}{1-e^{-\ii\left(\theta,\beta^{(p)}\right)}}\prod_{\alpha\in\tilde{\Delta}^{c}_+}\frac{1}{1-e^{-\ii\left(\theta,\alpha\right)}}
	\no
	&\times&
	\left.
		\prod_{\alpha\in\tilde{\Delta}^{n}_+\setminus\{\beta^{(p)}\}}\frac{1}{1+e^{-\ii\left(\theta,\alpha\right)}}
	\right] + \mathcal{O}\left(e^{\ii\left(\gamma,\theta\right)}\right),
\label{eqn43}
\eqne
where we again have excluded a factor proportional to $q^{(\ldots)}$.

Consider now the combination
\eqnb
\int d\theta\left[\chi^1_\mu(\theta)\chi_{\tilde{\mu}}^{\mf{so}(2p,1)}\left(\theta\right)-\Sigma^1_\mu(\theta)\Sigma_{\tilde{\mu}}^{\mf{so}(2p,1)}\left(\theta\right)\right],\label{eq54}
\eqne
where $\chi^1_\mu(\theta)$ and $\Sigma^1_\mu(\theta)$ are the grade zero part of the equations given in eqs.\ (\ref{char1}) and (\ref{sign1}), respectively. This is twice the number of non-unitary states at each grade. Therefore, to have unitarity, this integral has to be zero. Consider two different cases. The first case is when $\tilde\mu$ is not an integer. Then eq.\ (\ref{eq54}) reads
\eqnb
\left(-1\right)^{\left[\tilde{\mu}^{p}\right]}2\sum_{m=0}^{\infty}\sum_{n=0}^{\infty}\int d\theta e^{\ii\left(\mu+\tilde{\mu}+2\tilde{\rho}-\left(\left[\tilde{\mu}^{p}\right]+2+m+n\right)\beta^{(p)},\theta\right)},
\label{eq45}
\eqne
The second case is when $\tilde{\mu}^{p}$ is an even integer, then (\ref{eq54}) reads 
\eqnb
-2\sum_{m=0}^{\infty}\int d\theta e^{\ii\left(\mu+\tilde{\mu}+2\tilde{\rho}-\left(\tilde{\mu}^{p}+1+m\right)\beta^{(p)},\theta\right)}.
\label{eq46}
\eqne
Therefore, for the two different cases, we get non-zero result if
\eqnb
\mu+\tilde{\mu}+2\tilde{\rho}-\left(\left[\tilde{\mu}^{p}\right]+2+m+n\right)\beta^{(p)}
	&=&
		0
	\\
\mu+\tilde{\mu}+2\tilde{\rho}-\left(\tilde{\mu}^{p}+1+m\right)\beta^{(p)}
	&=&
		0.
\eqne
These solutions do not give antidominant weights since for the two cases
\eqnb
\mu^p+\mu^{p+1} &=& \left(\left[\tilde{\mu}^p\right] + 1 - \tilde{\mu}^p\right) + \left[\tilde{\mu}^p\right] + 1 + 2\left(m + n\right) > 0\no
\mu^p+\mu^{p+1} &=& \left(\left[\tilde{\mu}^p\right] - \tilde{\mu}^p\right) + \left[\tilde{\mu}^p\right] + 2m \geq 0.
\eqne
Hence, to be able to study conditions coming from the unitarity one has to consider other states, which in turn implies that one has to consider different cases depending on the values of the different components of the auxiliary highest weight. This will be even more complicated than the considerations in \cite{Bjornsson:2009dp} of the case $\mf{g}=\mf{su}(p,1)$. Furthermore, the study of the necessary conditions of $\mf{g}=\mf{su}(p,1)$ in \cite{Bjornsson:2009dp} was based on an iterative method, which here means that one has to show that $\tilde{\mu}^j \in \Z$ for $j>i$ in order to show that $\tilde{\mu}^i$ is an integer. 

\sect{Concluding remarks}
We have been able to prove unitarity for our models for antidominant highest weight representations. In addition, for the auxiliary we needed to impose the requirement of oddness of the last component of the highest weight. This requirement followed by the assuming that the integrand of eq.\ (\ref{eq328}) should vanish. The extra requirement is certainly not very appealing, as one would expect that it is not respected when one considers interactions. However, the condition may not be a necessary one. The necessary conditions arise only after performing the integration i.e.\ the integrand can be allowed to be non-zero. It is certainly an important problem, alas quite difficult, to study whether or not the extra condition is really needed. At this point we have no indications to what we may expect of such an investigation. 

Our construction here  works, as pointed out above, for an odd dimensional space-time. Basically, this is because we have only considered discrete representations using a Cartan-Weyl basis of the algebra. Certainly there exists a solution to the embedding problem using another basis. However, the analysis of unitarity will then have to proceed along a quite different lines. \vspace{1cm}

\noindent
{\bf{Acknowledgements}}
We would like to acknowledge the collaboration with Jens Fjelstad during the initial phase of this work. We also like to thank him for stimulating discussions during the progress of the paper. We would also like to thank Arkady Tseytlin for his remarks regarding the geometry of different gaugings. J.B.\ would like to thank the theory group at Karlstad University for the hospitality during the completion of this work. The work by J.B.\ is supported by the Swedish Research Council under project no.\ 623-2008-7048. S.H.\ is partially supported by the Swedish Research Council under project no.\ 621-2008-4129.

\newpage


\appendix


\sect{Some properties of $\mf{so}(4,2)$ and the subalgebra $\mf{so}(4,1)$}

In this appendix we will derive some properties of $\mf{so}(2p,2)$ and its subalgebra $\mf{so}(2p,1)$. This is the simplest non-trivial example of finite-dimensional algebras connected to the theories considered in the paper. We have included this discussion in some detail, since we have found nothing in the literature discussing certain issues of representations for non-compact algebras and possible embeddings of subalgebras. Since $\mf{so}(4,2)$ is a subalgebra $\mf{so}(2p,2)$ for $p>2$, some of the results presented here are also relevant for the more general case.


\subsection{The algebra $D_3$}

We first collect some formulas for $D_3$. The Cartan matrix is
\eqnb
A^{ij}  = 
    \left[
    \begin{array}{ccc}
      2     &-1    &-1    \\
      -1    &2     &0     \\
      -1    &0     &2     \\
    \end{array}
     \right],
\eqne
and the quadratic form is
\eqnb
G_{ij}  = 
    \left[
    \begin{array}{ccc}
      1     &1/2    & 1/2    \\
     1/2    &3/4     &1/4    \\
     1/2   &1/4     &3/4     \\
    \end{array}
     \right].
\eqne
The quadratic form is $G^{ij}=A^{ij}$, which follows from the fact that the Cartan matrix is symmetric.

There are three simple roots $\al^{(1)}$, $\al^{(2)}$ and $\al^{(3)}$. They all have length squared equal to two. The other positive roots are then $\al^{(1)i}+\al^{(2)i}=(1,1,-1)$, $\al^{(1)i}+\al^{(3)i}=(1,-1,1)$ and $\al^{(1)i}+\al^{(2)i}+\al^{(3)i}=(0,1,1)$. We also have $\al^{(1)}_i=(1,0,0)$, $\al^{(2)}_i=(0,1,0)$ and $\al^{(3)}_i=(0,0,1)$. One has $\rho^i\equiv \frac{1}{2}\sum_{\al>0}\al=(1,1,1)$. Denote the generators by $E^{\pm 1} \equiv E^{\pm \al^{(1)}}$, $E^{\pm 2}\equiv E^{\pm \al^{(2)}}$, $E^{\pm 3}\equiv E^{\pm \al^{(3)}}$, $E^{\pm 12}\equiv E^{\pm (\al^{(1)}+\al^{(2)})}$, $E^{\pm 13}\equiv E^{\pm (\al^{(1)}+\al^{(3)})}$ and $E^{\pm 123}$. The Cartan subalgebra is generated by $H^1$, $H^2$, and $H^3$. 


\subsection{Different realizations of the real form $\mf{so}(4,2)$}

In this section we will construct the different realizations of the real form of $\mf{so}(4,2)$ where the Cartan subalgebra is assumed to be compact. In order to find the real form from the complex Lie algebra, one needs to specify the Hermite conjugation properties. First, we take all elements of the Cartan algebra to be Hermitean, $(H^i)^\dagger =H^i$ for $i=1,2,3$. This corresponds to choosing them to be belong to the compact subalgebra of $\mf{so}(4,2)$ (the maximal compact subalgebra is given by $\mf{so}(4)\oplus \mf{u}(1)$).
 
Next, we assume that all generators $E^\al$ are either compact or non-compact implying that $(E^\al)^\dagger=\pm E^{-\al}$, where the plus-sign is for a compact case. Since $\mf{so}(4,2)$ has seven compact and eight non-compact directions, and we have three compact Cartan generators, the remaining generators group into four compact and eight non-compact generators. Thus, we have four compact generators $E^{\pm \al}$, $E^{\pm \be}$. The remaining four generators are non-compact. It is easy to see that the choice of realization of the real form induces a $\Z_2$ grading of the algebra. Let us now treat all possibilities systematically. We will, for simplicity, call $\alpha$ a (non-) compact root if $E^\alpha$ is a (non-) compact generator. 
\vs{3mm}
\\
\underline{Case I: $\al^{(1)}$ compact}\vs{5mm}
\\
If either $\al^{(2)}$ or $\al^{(3)}$ are compact, then the commutator implies that $\al^{(1)}+\al^{(2)}$ or $\al^{(1)}+\al^{(3)}$ are compact. This gives too many compact roots. Thus, $\al_2$ and $\al_3$ are non-compact. This implies that $\al^{(1)}+\al^{(2)}$ and $\al^{(1)}+\al^{(3)}$ are non-compact, and $\al^{(1)}+\al^{(2)}+\al^{(3)}$ is compact. This case we will denote as a realization of the real form of type $A$, or type $A$ real form for short.
\vs{5mm}
\\
\underline{Case II: $\al^{(1)}$ non-compact}\vs{5mm}
\\
Here we will have to treat several possibilities.
\\
{\bf (a)} $\al^{(2)}$ and $\al^{(3)}$ are compact. Then $\al^{(1)}+\al^{(2)}$, $\al^{(1)}+\al^{(3)}$ and $\al^{(1)}+\al^{(2)}+\al^{(3)}$ are non-compact. This case we will call a real form of type $B$.
\\
{\bf (b)} One of $\al^{(2)}$ and $\al^{(3)}$ is non-compact. Then, one of $\al^{(1)}+\al^{(2)}$ and $\al^{(1)}+\al^{(3)}$ is non-compact and $\al^{(1)}+\al^{(2)}+\al^{(3)}$ compact. This gives three compact roots, one too many.
\\
{\bf (c)} $\al^{(2)}$ and $\al^{(3)}$ are non-compact. This implies that both $\al^{(1)}+\al^{(2)}$ and $\al^{(1)}+\al^{(3)}$ are compact, while $\al^{(1)}+\al^{(2)}+\al^{(3)}$ is non-compact. This case we will call a real form of type $C$. 
  
This exhausts all possibilities and we have in total found three different types of real forms. These possibilities are not completely independent. By performing Weyl transformations one may rotate the different roots into each other. For example, performing a Weyl rotation generated by $\al^{(1)}$, one easily shows that starting from an $A$-type real form one will again get an $A$-type real form. On the other hand for a $B$-type real form it is transformed into a $C$-type real form and vice versa. Exactly the same result is found if we use $(\al^{(1)}+\al^{(2)}+\al^{(3)})$ as a generator of the Weyl transformations. The roots $\al^{(2)}$ and $\al^{(3)}$ on the other hand will change types as follows: $A\rightarrow C$, $B\rightarrow B$ and $C\rightarrow A$. Finally, $(\al^{(1)}+\al^{(2)})$ and $(\al^{(1)}+\al^{(3)})$ will give $A\rightarrow B$, $B\rightarrow A$ and $C\rightarrow C$. Let us write three examples of explicit transformations that relate the different types.
\vs{5mm}
\\
\underline{$A\rightarrow B$ }
\eqnb
\tilde{H^1}&=&H^2\no
\tilde{H^2}&=&H^1\no
\tilde{H^3}&=&-(H^1+H^2+H^3)\no
\tilde{E}^{\pm 1}&=&E^{\pm 2}\no
\tilde{E}^{\pm 2}&=&E^{\pm 1}\no
\tilde{E}^{\pm 3}&=&E^{\mp 123}\no
\tilde{E}^{\pm 12}&=&-E^{\pm 12}\no
\tilde{E}^{\pm 13}&=&-E^{\mp 13}\no
\tilde{E}^{\pm 123}&=&E^{\mp 3}\label{trans1}
\eqne
\underline{$B\rightarrow C$ }
\eqnb
\tilde{H^1}&=&H^1\no
\tilde{H^2}&=&-(H^1+H^2)\no
\tilde{H^3}&=&-(H^1+H^3)\no
\tilde{E}^{\pm 1}&=&E^{\pm 1}\no
\tilde{E}^{\pm 2}&=&E^{\mp 12}\no
\tilde{E}^{\pm 3}&=&E^{\mp 13}\no
\tilde{E}^{\pm 12}&=&-E^{\mp 2}\no
\tilde{E}^{\pm 13}&=&-E^{\mp 3}\no
\tilde{E}^{\pm 123}&=&E^{\mp 123}\label{trans2}
\eqne
\underline{$A\rightarrow C$ }
\eqnb
\tilde{H^1}&=&H^2\no
\tilde{H^2}&=&-(H^1+H^2)\no
\tilde{H^3}&=&-(H^1+H^3)\no
\tilde{E}^{\pm 1}&=&E^{\pm 2}\no
\tilde{E}^{\pm 2}&=&E^{\mp 12}\no
\tilde{E}^{\pm 3}&=&-E^{\mp 13}\no
\tilde{E}^{\pm 12}&=&-E^{\mp 1}\no
\tilde{E}^{\pm 13}&=&-E^{\pm 123}\no
\tilde{E}^{\pm 123}&=&E^{\mp 3}\label{trans3}
\eqne

  
\subsection{Highest and lowest weight representations of $\mf{so}(4,2)$}
   
We proceed by investigating different possibilities of highest and lowest weight representations. For $\mf{sl}(2)$ the only possibilities are either a highest weight representation or a lowest weight representation. However, for other algebras there are more possibilities. These are given by examining all possible sets of roots $\Delta_1$ such that the following conditions are consistent
\eqnb
E^\be\left|\mu\right> =0,\hspace{3mm} \be\in \Delta_1.
\eqne
We will now show that there are exactly 24 different possibilities. It is obvious that if we have one possibility $\Delta_1$, then there exists another one given by changing all the signs of the roots in $\Delta_1$. Let us denote this as the dual set $\bar{\Delta}_1$ Thus, we can  concentrate on the 12 remaining possibilities by assuming that $\al^{(1)}\in \Delta_1$.
 
 The simplest case is given by assuming $\al^{(2)}$ and $\al^{(3)}$ are also in $\Delta_1$. Then all positive roots are in $\Delta_1$. This is the highest weight representation. The dual representation is the lowest weight representation.

Next, assume $\al^{(2)}$ and $-\al^{(3)}$ are in $\Delta_1$. Taking the commutator of $E ^1$ and $E^2$ we find that $\al^{(1)}+\al^{(2)}\in \Delta_1$. For $\pm(\al^{(1)}+\al^{(3)})$ we have two possibilities either  $\al^{(1)}+\al^{(3)}\in \Delta_1$ or $-(\al^{(1)}+\al^{(3)})\in \Delta_1$. In the first case the commutator between $E^2$ and $E^{13}$ implies that  $\al^{(1)}+\al^{(2)}+\al^{(3)}\in \Delta_1$. In the second case we can consistently take either $\al^{(1)}+\al^{(2)}+\al^{(3)}\in \Delta_1$ or $-(\al^{(1)}+\al^{(2)}+\al^{(3)})\in \Delta_1$. In total, this gives three different possibilities.

The third possibility is to have $-\al^{(2)}$ and $\al^{(3)}$ are in $\Delta_1$. The discussion is exactly the same as the one above interchanging the roles of $\pm\al^{(2)}$ and $\pm\al^{(3)}$. This gives three additional cases.

The fourth possibility is to have $-\al^{(2)}$ and $-\al^{(3)}$ in $\Delta_1$. Then we may choose either $\al^{(1)}+\al^{(2)}\in \Delta_1$ or $-(\al^{(1)}+\al^{(2)})\in \Delta_1$. For each of these alternatives we can take either $\al^{(1)}+\al^{(3)}\in \Delta_1$ or $-(\al^{(1)}+\al^{(3)})\in \Delta_1$. Checking the different commutators one finds in total five different consistent possibilities (given below). Summarizing one finds the twelve cases to be:
\vsv
\\
\underline{Case (a)}\\ $\Delta_1$ consists of all positive roots i.e. we have a highest weight representation.
\vsv
\\
\underline{Case (b)}\\ $\al^{(1)}, \al^{(2)}, -\al^{(3)}, \al^{(1)}+\al^{(2)}, \al^{(1)}+\al^{(3)},\al^{(1)}+\al^{(2)}+\al^{(3)} \in \Delta_1$.
\vsv
\\
\underline{Case (c)}\\ $\al^{(1)}, \al^{(2)}, -\al^{(3)}, \al^{(1)}+\al^{(2)}, -(\al^{(1)}+\al^{(3)}),\al^{(1)}+\al^{(2)}+\al^{(3)} \in \Delta_1$.
\vsv
\\
\underline{Case (d)}\\ $\al^{(1)}, \al^{(2)}, -\al^{(3)}, \al^{(1)}+\al^{(2)}, -(\al^{(1)}+\al^{(3)}),-(\al^{(1)}+\al^{(2)}+\al^{(3)}) \in \Delta_1$.
\vsv
\\
\underline{Case (e)}\\ $\al^{(1)}, -\al^{(2)}, \al^{(3)}, \al^{(1)}+\al^{(2)}, \al^{(1)}+\al^{(3)},\al^{(1)}+\al^{(2)}+\al^{(3)} \in \Delta_1$.
\vsv
\\
\underline{Case (f)}\\ $\al^{(1)}, -\al^{(2)}, \al^{(3)}, -(\al^{(1)}+\al^{(2)}), \al^{(1)}+\al^{(3)},\al^{(1)}+\al^{(2)}+\al^{(3)} \in \Delta_1$.
\vsv
\\ 
\underline{Case (g)}\\ $\al^{(1)}, -\al^{(2)}, \al^{(3)}, -(\al^{(1)}+\al^{(2)}), \al^{(1)}+\al^{(3)},-(\al^{(1)}+\al^{(2)}+\al^{(3)}) \in \Delta_1$.
\vsv
\\ 
\underline{Case (h)}\\ $\al^{(1)}, -\al^{(2)}, -\al^{(3)}, \al^{(1)}+\al^{(2)}, \al^{(1)}+\al^{(3)},\al^{(1)}+\al^{(2)}+\al^{(3)} \in \Delta_1$.
\vsv
\\ 
\underline{Case (i)}\\ $\al^{(1)}, -\al^{(2)}, -\al^{(3)}, \al^{(1)}+\al^{(2)}, \al^{(1)}+\al^{(3)},-(\al^{(1)}+\al^{(2)}+\al^{(3)}) \in \Delta_1$.
\vsv
\\ 
\underline{Case (j)}\\ $\al^{(1)}, -\al^{(2)}, -\al^{(3)}, \al^{(1)}+\al^{(2)}, -(\al^{(1)}+\al^{(3)}),-(\al^{(1)}+\al^{(2)}+\al^{(3)}) \in \Delta_1$.
\vsv
\\ 
\underline{Case (k)}\\ $\al^{(1)}, -\al^{(2)}, -\al^{(3)}, -(\al^{(1)}+\al^{(2)}), \al^{(1)}+\al^{(3)},-(\al^{(1)}+\al^{(2)}+\al^{(3)}) \in \Delta_1$.
\vsv
\\ 
\underline{Case (l)}\vs{-2mm}\eqnb\hs{-19mm}\al^{(1)}, -\al^{(2)}, -\al^{(3)}, -(\al^{(1)}+\al^{(2)}), -(\al^{(1)}+\al^{(3)}),-(\al^{(1)}+\al^{(2)}+\al^{(3)}) \in \Delta_1.
\label{weightcomb}
\eqne
All dual cases are found by taking $\bar{\Delta}_1=-\Delta_1.$

The different mixtures of highest and lowest weight conditions found above may be understood in a simple way. Looking at the transformations (\ref{trans1})-(\ref{trans3}), we see that if we define, for example, a highest weight representation with respect to the untilded generators, then with respect to the tilded operators, the representation is not a highest weight representation. Instead it corresponds to case (d) above. Thus, using Weyl transformations one may connect different possibilities (a)-(k). In fact, all cases can be derived in this way, which can be proven in a straightforward way by simply performing all 24 Weyl transformations of the Weyl group of $D_3$. We have no simple argument why this should be true. It is of course so that every Weyl transformation yields a consistent set of conditions transforming from e.g. a highest weight representation. However, that the converse is also true is not,
as far as we can tell, evident.

The fact that all the different combinations of highest and lowest weight conditions defining a discrete representation of $\mf{so}(4,2)$ are connected through the inner automorphisms of the algebra generated by the Weyl group means that we can arbitrarily choose one of the 24 possibilities. Finding all possible unitary representations for this choice will imply that one has found all possible unitary representations for all the other cases as well. Thus, we could, for example, choose to study only the highest weight representations.

However,  once one has chosen a particular set of conditions to study, one has fixed the symmetry of the Weyl group and one is no longer free to use the Weyl group to connect the different types of real forms. Hence, one has to study all three types. They will, as we will see, not be equivalent possibilities from the point of unitarity.

There is another option, namely to use the Weyl group to fix the type of real form. Then this leaves a some part of the Weyl group, but not all of it, to relate the different combinations in eq.\ (\ref{weightcomb}). One reason to fix the type of real form is that when we look at embeddings of the subalgebra $\mf{so}(4,1)$ into $\mf{so}(4,2)$ then the different real forms admit different embeddings, as we will discuss below. The most natural embedding from the point of view of finding the coset space is the type $A$ real form.

Let us now show that the different types of real forms are not equivalent as far as unitarity is concerned, if we fix the type of representation completely. We choose to fix the type of representation to be of highest weight i.e.\ option (a) above in eq.\ (\ref{weightcomb}). We will now show that both type $A$ and $C$ real forms have no unitary highest weight representations. For type $A$ real form $E^{\pm 1}$, $E^{\pm 123}$ and $H^i$, $i=1,2,3$ are compact generators and all the others are non-compact. Since $E^{\pm \be}$, with $\be$ being a positive root, and a corresponding element in the Cartan subalgebra need to have unitary representations of either $\mf{su}(2)$ for the compact case, or $\mf{su}(1,1)$ for the non-compact case. Then we must have that the highest weight $\mu$ of a highest weight state, has a positive (or zero) component in the compact direction and a negative component in the non-compact direction. This implies $\mu^1\geq 0$, $\mu^2\leq 0$, $\mu^3\leq 0$, $\mu^1+\mu^2\leq 0$, $\mu^1+\mu^3\leq 0$, and $\mu^1+\mu^2+\mu^3\geq 0$. Adding the fourth and fifth conditions we get $2\mu^1+\mu^2+\mu^3\leq 0$, which is not consistent with what is implied by the first and last conditions. Thus, there are no unitary highest weight representations.

For the real form of type $C$ the argument is analogous. Here $E^{\pm 12}$ and $E^{\pm 13}$ are compact generators together with the Cartan generators. Then using the same argument as above we have $\mu^i\leq 0$, $i=1,2,3$, $\mu^1+\mu^2\geq 0$, $\mu^1+\mu^3\geq 0$, and $\mu^1+\mu^2+\mu^3\leq 0$. Clearly, these requirements are incompatible. Thus, there are no unitary highest weight representations for this type of real form, either. 

For the remaining real form of type $B$ we have $E^{\pm 2}$ and $E^{\pm 3}$ as compact generators. Then we get the following conditions, $\mu^1\leq 0$, $\mu^2\geq 0$, $\mu^3\geq 0$, $\mu^1+\mu^2\leq 0$,$\mu^1+\mu^3\leq 0$, and $\mu^1+\mu^2+\mu^3\leq 0$, which are equivalent to $\mu^1\leq 0$, $\mu^2\geq 0$, $\mu^3\geq 0$ and $\mu^1+\mu^2+\mu^3\leq 0$. These conditions are compatible with each other, and one does have unitary highest weight representations, as proven by Jakobsen \cite{Jakobsen:1983} and Enright {\it et al} \cite{Enright:1983}.

Let us summarize our results for $\mf{so}(4,2)$ in a lemma.

\begin{lemma}
There are precisely three inequivalent realizations of the real form $\mf{so}(4,2)$ in $D_3$ where the Cartan subalgebra is compact. Among the three types, only one, denoted type $B$ here, admits unitary highest weight representations.
\end{lemma}

Note that since we do not have to require unitary highest weight representations of $\mf{so}(4,2)$ in our construction, we are not forced to choose the type $B$ real form. 


\subsection{$\mf{so}(4,1)$ as subalgebra of $\mf{so}(4,2)$} 

The Cartan matrix for $B_2$ is 
\eqnb
A^{ij}  = 
    \left[
    \begin{array}{cc}
      2     &-2       \\
      -1    &2        \\
    \end{array}
     \right],
\eqne
and the quadratic forms 
\eqnb
G_{ij}  = 
    \left[
    \begin{array}{cc}
      1     &1/2       \\
      1/2   & 1/2       \\
    \end{array}
     \right] \hs{8mm}
 G^{ij}  = 
    \left[
    \begin{array}{cc}
      2     &-2       \\
      -2   & 4       \\
    \end{array}
     \right] .
\eqne

The two simple roots of $B_2$ we denote by $\be^{(1)}$ and $\be^{(2)}$. $\be^{(1)}$ is a long root, $(\be^{(1)},\be^{(1)})=2$, and $\be^{(2)}$ is a short root, $(\be^{(2)},\be^{(2)})=1$. Furthermore, $(\be^{(1)},\be^{(2)})=-1$. We have explicitely $\be^{(1)i}=(2,-2)$, $\be^{(2)i}=(-1,2)$, $\be^{(1)}_i=(1,0)$ and $\be^{(2)}_i=(0,1/2)$. $\rho\equiv \frac{1}{2}\sum_{\alpha>0}\alpha= \frac{3}{2}\be^{(1)}+2\be^{(2)}$. Denote the generators by $F^{\pm 1} \equiv F^{\pm \be^{(1)}}$, $F^{\pm 2}\equiv F^{\pm \be^{(2)}}$, $F^{\pm 12}\equiv F^{\pm (\be^{(1)}+\be^{(2)})}$, $F^{\pm 122}\equiv F^{\pm (\be^{(1)}+2\be^{(2)})}$. The Cartan subalgebra is generated by $K^1$ and $K^2$.

Let us now investigate what type of realizations of $\mf{so}(4,1)$ may occur. $\mf{so}(4,1)$ has six compact generators. For $\mf{so}(4,2)$ we found three different types. We will show that for $\mf{so}(4,1)$ there is only one type that one can construct assuming the Cartan subalgebra is compact.
\vsv
\\
(i) Assume first that $F^{1}$ is compact. Then $F^{2}$ has to be non-compact, which implies that $F^{12}$ is non-compact and $F^{122}$ compact. This is a possible solution.
\vsv
\\
(ii) Assume $F^{1}$ to be non-compact. If $F^{2}$ is compact, then $F^{12}$ and $F^{122}$ are non-compact, which is not a correct solution. If $F^{2}$ is non-compact, then $F^{12}$ is compact and $F^{122}$ is non-compact, which is not a correct solution, either. This concludes the proof.
\vsv

Thus, we have found that the only possibility is to have $F^{\pm 1}$, $F^{\pm 122}$ and $K^i$ to be compact, under the assumption that the Cartan algebra is compact.

We should now like to investigate in what way $\mf{so}(4,1)$ may be embedded into $\mf{so}(4,2)$. As it will turn out, this will depend on what realization of the real form of $D_3$ we choose. In particular, we will show that if we would like to have an embedding which preserves the triangular structure of $D_3$, i.e.\ positive (negative) root generators of $B_2$ is formed from linear combinations of positive (negative) root generators of $D_3$, and in addition, the Cartan subalgebra of $B_2$ is a subalgebra of $D_3$, then this will only be possible for the type $A$ real form of the latter algebra. We will call an embedding that preserves the triangular structure to be a {\it regular} embedding.

\begin{lemma}
The only regular embedding of $\mf{so}(4,1)$ in $\mf{so}(4,2)$, where the Cartan subalgebra is compact, appears for the type $A$ real form.
\end{lemma}

\paragraph{Proof.} We begin by assuming the type $A$ real form. This means that $\al^{(1)}$ and $\al^{(1)}+\al^{(2)}+\al^{(3)}$ correspond to compact generators, while the rest of the roots correspond to non-compact ones. As we have established that there is only one real form of $B_2$ with the Cartan subalgebra being compact, we must for a regular embedding take the compact generators of $\mf{so}(4,1)$ i.e.\ $F^{\pm 1}$ and $F^{\pm 122}$ to be linear combinations of $E^{\pm 1}$ and $E^{\pm 123}$. A general ansatz is to take $F^{1}=aE^{1}+bE^{\pm 123}$ and $F^{122}=cE^{1}+dE^{123}$ for constants $a,b,c,d$. Taking the Hermitean conjugation we have $F^{-1}=a^\ast E^{-1}+b^\ast E^{-123}$ and $F^{-122}=c^\ast E^{-1}+d^\ast E^{-123}$. From $[F^{1},F^{-122}]=0$ we get $ac^\ast=bd^\ast=0$. We have two solutions either $b=c=0$ or $a=d=0$. Let us choose $b=c=0$ and $a=d=1$ (for simplicity). The other solution, $b=c=1$ and $a=d=0$ is easily seen not to lead a possible solution following the steps below.

To find the other generators one can make a general ansatz for $F^2$ using the non-compact generators of $\mf{so}(4,2)$. $F^{2}=pE^{2}+qE^{3}+rE^{12}+sE^{13}$. Then we have from $[F^1,F^2]$, $F^{12}=[F^1,F^2]=pE^{12}+qE^{13}$ and $F^{122}=\frac{1}{2}[F^2,F^{12}]=(p+q)E^{123}$. Using $F^{-1}=E^{-1}$ and $[F^{-1}, F^{12}]=F^{2}$, we see that $r=s=0$. We can, furthermore, set $p=q=1$. We have then found the following complete embedding
\eqnb
F^{\pm 1}&=& E^{\pm 1} \no
F^{\pm 2}&=& E^{\pm 2}+E^{\pm 3}\no
F^{\pm 12}&=& E^{\pm 12}+E^{\pm 13}\no
F^{\pm 122} &=& E^{\pm 123}\no
K^1&=& H^1\no
K^2&=& H^2+H^3.\label{embedding2}
\eqne

Let us now look at type $B$ real form. Then $E^{\pm 2}$ and $E^{\pm 3}$ are compact, while the rest are non-compact. We have a general ansatz $F^1=aE^2+bE^3$, $F^{122}=cE^2+dE^3$ and $F^2=pE^{1}+qE^{12}+rE^{13}+sE^{123}$. Using again the commutator $[F^{1},F^{-122}]=0$, we arrive at the same conditions as above, $b=c=0$ or $a=d=0$. Since the algebra $D_3$ is symmetric under interchange $E^2\leftrightarrow E^3$, we can, without loss of generality, choose $b=c=0$ and $a=d=1$. Then $F^{12}=[F^1,F^2]=-pE^{12}+rE^{123}$ and $F^{122}=\frac{1}{2}[F^2,F^{12}]=0$. Hence, it is not possible to find a regular solution to the embedding problem.   		      
   
For the type $C$ real form we have $E^{\pm 12}$ and $E^{\pm 13}$ as compact generators. Using the same type of ansatz as before we find that we can without loss of generality take $F^1=E^{12}$ and $F^{122}=E^{13}$. We have, furthermore, the ansatz $F^2=pE^{1}+qE^{2}+rE^{3}+sE^{123}$. This implies $F^{12}=[F^1,F^2]=rE^{123}$ and $F^{122}=\frac{1}{2}[F^2,F^{12}]=0$. Again, we see that we have no regular solution.\ $\Box$

There do exist, however, other embeddings than regular ones. We can use the automorphisms of the algebra discussed previously to find such possibilities. For example, using the transformations eq.\ (\ref{trans1}), we can get an embedding for type $B$ real form from the one found for type $A$ real form eq.\ (\ref{embedding2}). We have explicitely,
\eqnb
F^{\pm 1}&=& E^{\pm 2} \no
F^{\pm 2}&=& E^{\pm 1}+E^{\mp 123}\no
F^{\pm 12}&=& -E^{\pm 12}-E^{\mp 13}\no
F^{\pm 122} &=& E^{\mp 3}\no
K^1&=& H^2\no
K^2&=& -H^2-H^3.\label{embeddingB}
\eqne  
Likewise, we get an embedding for the type $C$ real form,
\eqnb
F^{\pm 1}&=& -E^{\mp 12} \no
F^{\pm 2}&=& E^{\pm 1}+E^{\pm 123}\no
F^{\pm 12}&=& -E^{\mp 2}-E^{\pm 3}\no
F^{\pm 122} &=& -E^{\pm 13}\no
K^1&=& -H^1-H^2\no
K^2&=& H^1.\label{embeddingC}
\eqne

Let us now discuss the signature function of $\mf{so}(4,1)$. We will only consider the horizontal algebra, although the affine case is easily treated analogously. We use the real form that can be regularly embedded into $\mf{so}(4,2)$, i.e. $\beta^{(1)}$ and $\beta^{(1)}+2\beta^{(2)}$ correspond to compact generators while $\beta^{(2)}$ and $\beta^{(1)}+\beta^{(2)}$ correspond to non-compact ones. We now consider representations of $\mf{so}(4,1)$ that are appropriate for the auxiliary sector. But we will denote the generators by $E^\al$ (without tilde, for simplicity). The state-space is generated by a highest weight state $\left|\mu\right>$. Then the state-space is spanned by states of the form
\eqnb
\left|\lambda\right>\equiv \prod_{m=1}^{p}E^{\al_m}\left|\mu\right>.
\eqne
Here $\lambda=\mu-n_1\bi-n_2\bii$, where $n_1,n_2$ are non-negative integers.
\begin{lemma}
Let $\mu$ be a dominant integral weight. The signature of the norm of the state $\left|\la\right>$ is given by $(-1)^{n_2}$.
\label{lemmai}
\end{lemma}
\paragraph{Proof.}
The state $\left|\la\right>$ can be written in the form
\eqnb
\left|\la\right>= \sum_{\mathrm{permutations}}\left[C_{\be_1\be_2\ldots \be_{n_1+n_2}} 
E^{\beta_1}E^{\beta_2}\ldots E^{\be_{n_1+n_2}}\right]\left|\mu\right>
\eqne
where $\be_i$, $i=1,\ldots , n_1+n_2$, are either $\bi$ ($n_1$ factors) or $\bii$ ($n_2$ factors). Taking the Hermitian conjugate of this state yields the sign factor $(-1)^{n_2}$ from the non-compact generators. Then the norm is given by this sign factor multiplied by a non-negative constant. This follows from the fact that an irreducible representation of integral dominant highest weight is unitary for $\mf{so}(5)$, and the sign factor is the only difference when going to the $\mf{so}(4,1)$ case. If the state is outside the irreducible representation then it has zero norm.
$\Box$

The character for the irreducible highest weight representation is given by Weyl's character formula.
\eqnb
\chi_\mu(\theta_1,\theta_2)=\sum_{w\in W_{B_2}}\sign(w)\frac{e^{i\left(w(\mu+\rho)-\rho,\theta\right)}}{\prod_{\al\in \Delta^+}\left(1-e^{-i\left(\al,\theta\right)}\right)}.
\eqne
The signature function is given by
\eqnb
\Sigma_\mu(\theta_1,\theta_2)=\sum_{w\in W_{B_2}}\sign(w)
\frac{(-1)^{2\left(w(\mu+\rho)-\mu-\rho,\Lambda^{(2)}\right)}
e^{i\left(w(\mu+\rho)-\rho,\theta\right)}}
{\prod_{\al\in \Delta^+_c}\left(1-e^{-i\al\cdot\theta}\right)
\prod_{\al\in \Delta^+_n}\left(1+e^{-i\al\cdot\theta}\right)}
\eqne
where $\Lambda^{(2)}$ is a fundamental weight (corresponding to $\bii$). The signature function is proven using Lemma \ref{lemma5}.

\newpage


\end{document}